\documentclass[twocolumn]{aastex631}

\begin{document}

\title{SN 2023ixf in M101: physical parameters from bolometric light curve modeling}

\correspondingauthor{Jozsef Vinko}
\email{vinko.jozsef@csfk.hun-ren.hu}

\author[0000-0001-8764-7832]{J\'ozsef Vink\'o}
\affiliation{HUN-REN CSFK Konkoly Observatory, MTA Centre of Excellence, 
Konkoly Thege M. \'ut 15-17, Budapest, 1121, Hungary}
\affiliation{Department of Experimental Physics, Institute of Physics, University of Szeged, D\'om t\'er 9, Szeged, 6720 Hungary}
\affiliation{ELTE E\"otv\"os Lor\'and University, Institute of Physics and Astronomy,  
P\'azm\'any P\'eter s\'et\'any 1A, Budapest 1117, Hungary }
\affiliation{Department of Astronomy, University of Texas at Austin,
2515 Speedway, Stop C1400, Austin, TX, 78712-1205, USA}

\author[0009-0006-4162-5447]{Zs\'ofia R\'eka Bodola}
\affiliation{Department of Experimental Physics, Institute of Physics, University of Szeged, D\'om t\'er 9, Szeged, 6720 Hungary}

\author[0009-0001-3178-603X]{\'Akos G{\H o}d\'eny}
\affiliation{ELTE E\"otv\"os Lor\'and University, Institute of Physics and Astronomy,  
P\'azm\'any P\'eter s\'et\'any 1A, Budapest 1117, Hungary }

\author[0009-0000-0914-130X]{Szelina Fruzsina Cs\'ak}
\affiliation{ELTE E\"otv\"os Lor\'and University, Institute of Physics and Astronomy,  
P\'azm\'any P\'eter s\'et\'any 1A, Budapest 1117, Hungary }

\author[0000-0002-8770-6764]{R\'eka K\"onyves-T\'oth}
\affiliation{HUN-REN CSFK Konkoly Observatory, MTA Centre of Excellence, Konkoly Thege M. \'ut 15-17, Budapest, 1121, Hungary}
\affiliation{Department of Experimental Physics, Institute of Physics, University of Szeged, D\'om t\'er 9, Szeged, 6720 Hungary}

\author[0000-0002-9324-3903]{Andrea P. Nagy}
\affiliation{Department of Experimental Physics, Institute of Physics, University of Szeged, D\'om t\'er 9, Szeged, 6720 Hungary}

\author[0000-0003-4610-1117]{Tam\'as Szalai}
\affiliation{Department of Experimental Physics, Institute of Physics, University of Szeged, D\'om t\'er 9, Szeged, 6720 Hungary}

\author[0009-0000-9929-7518]{Dominik B\'anhidi}
\affiliation{Baja Observatory of University of Szeged, Szegedi \'ut, Kt. 766, Baja, 6500, Hungary}

\author[0000-0001-9061-2147]{Imre Barna B\'ir\'o}
\affiliation{Baja Observatory of University of Szeged, Szegedi \'ut, Kt. 766, Baja, 6500, Hungary}
\affiliation{HUN-REN-SZTE Stellar Astrophysics Research Group, Szegedi \'ut, Kt. 766, Baja, 6500, Hungary}

\author[0000-0002-8585-4544]{Attila B\'odi}
\affiliation{Department of Astrophysical Sciences, Princeton University Peyton Hall, 4 Ivy Lane, Princeton, NJ 08544, USA}
\affiliation{HUN-REN CSFK Konkoly Observatory, MTA Centre of Excellence, 
Konkoly Thege M. \'ut 15-17, Budapest, 1121, Hungary}

\author[0000-0001-6232-9352]{Zs\'ofia Bora}
\affiliation{ELTE E\"otv\"os Lor\'and University, Institute of Physics and Astronomy, 
P\'azm\'any P\'eter s\'et\'any 1A, Budapest 1117, Hungary }
\affiliation{HUN-REN CSFK Konkoly Observatory, MTA Centre of Excellence, 
Konkoly Thege M. \'ut 15-17, Budapest, 1121, Hungary}

\author{Istv\'an  Cs\'anyi}
\affiliation{Baja Observatory of University of Szeged, Szegedi \'ut, Kt. 766, Baja, 6500, Hungary}

\author[0000-0002-6497-8863]{Borb\'ala Cseh}
\affiliation{HUN-REN CSFK Konkoly Observatory, MTA Centre of Excellence, 
Konkoly Thege M. \'ut 15-17, Budapest, 1121, Hungary}
\affiliation{MTA-ELTE Lend{\"u}let "Momentum" Milky Way Research Group, Hungary}

\author[0009-0004-3558-8948]{Tibor Heged\"us}
\affiliation{Baja Observatory of University of Szeged, Szegedi \'ut, Kt. 766, Baja, 6500, Hungary}

\author[0009-0008-2052-8474]{\'Agoston Horti-D\'avid}
\affiliation{ELTE E\"otv\"os Lor\'and University, Institute of Physics and Astronomy, 
P\'azm\'any P\'eter s\'et\'any 1A, Budapest 1117, Hungary }
\affiliation{HUN-REN CSFK Konkoly Observatory, MTA Centre of Excellence, Konkoly Thege M. \'ut 15-17, Budapest, 1121, Hungary}

\author[0000-0001-5203-434X]{Andr\'as P\'eter Jo\'o}
\affiliation{ELTE E\"otv\"os Lor\'and University, Institute of Physics and Astronomy, 
P\'azm\'any P\'eter s\'et\'any 1A, Budapest 1117, Hungary }
\affiliation{HUN-REN CSFK Konkoly Observatory, MTA Centre of Excellence, 
Konkoly Thege M. \'ut 15-17, Budapest, 1121, Hungary}

\author[0000-0002-1663-0707]{Csilla Kalup}
\affiliation{ELTE E\"otv\"os Lor\'and University, Institute of Physics and Astronomy, 
P\'azm\'any P\'eter s\'et\'any 1A, Budapest 1117, Hungary }
\affiliation{HUN-REN CSFK Konkoly Observatory, MTA Centre of Excellence, 
Konkoly Thege M. \'ut 15-17, Budapest, 1121, Hungary}

\author[0000-0002-1792-546X]{Levente Kriskovics}
\affiliation{HUN-REN CSFK Konkoly Observatory, MTA Centre of Excellence, 
Konkoly Thege M. \'ut 15-17, Budapest, 1121, Hungary}

\author{Erika Mochnács}
\affiliation{Department of Experimental Physics, Institute of Physics, University of Szeged, D\'om t\'er 9, Szeged, 6720 Hungary}

\author[0000-0001-5449-2467]{Andr\'as P\'al}
\affiliation{HUN-REN CSFK Konkoly Observatory, MTA Centre of Excellence, 
Konkoly Thege M. \'ut 15-17, Budapest, 1121, Hungary}

\author[0000-0001-5573-8190]{Zsolt Reg\'aly}
\affiliation{HUN-REN CSFK Konkoly Observatory, MTA Centre of Excellence, 
Konkoly Thege M. \'ut 15-17, Budapest, 1121, Hungary}

\author[0000-0002-3658-2175]{B\'alint Seli}
\affiliation{ELTE E\"otv\"os Lor\'and University, Institute of Physics and Astronomy, 
P\'azm\'any P\'eter s\'et\'any 1A, Budapest 1117, Hungary }
\affiliation{HUN-REN CSFK Konkoly Observatory, MTA Centre of Excellence, 
Konkoly Thege M. \'ut 15-17, Budapest, 1121, Hungary}

\author[0000-0001-7806-2883]{\'Ad\'am S\'odor}
\affiliation{HUN-REN CSFK Konkoly Observatory, MTA Centre of Excellence, 
Konkoly Thege M. \'ut 15-17, Budapest, 1121, Hungary}

\author[0009-0007-1015-0327]{Norton Oliv\'er Szab\'o}
\affiliation{ELTE E\"otv\"os Lor\'and University, Institute of Physics and Astronomy, 
P\'azm\'any P\'eter s\'et\'any 1A, Budapest 1117, Hungary }
\affiliation{HUN-REN CSFK Konkoly Observatory, MTA Centre of Excellence, 
Konkoly Thege M. \'ut 15-17, Budapest, 1121, Hungary}

\author[0000-0002-1698-605X]{R\'obert Szak\'ats}
\affiliation{HUN-REN CSFK Konkoly Observatory, MTA Centre of Excellence, 
Konkoly Thege M. \'ut 15-17, Budapest, 1121, Hungary}

\author{P\'eter Sz\'ekely}
\affiliation{Department of Experimental Physics, Institute of Physics, University of Szeged, D\'om t\'er 9, Szeged, 6720 Hungary}
\affiliation{Semilab Inc., Prielle Korn\'elia utca 2, Budapest, 1117, Hungary}

\author{V\'azsony Varga}
\affiliation{ELTE E\"otv\"os Lor\'and University, Institute of Physics and Astronomy, 
P\'azm\'any P\'eter s\'et\'any 1A, Budapest 1117, Hungary }
\affiliation{HUN-REN CSFK Konkoly Observatory, MTA Centre of Excellence, 
Konkoly Thege M. \'ut 15-17, Budapest, 1121, Hungary}

\author[0000-0002-6471-8607]{Kriszti\'an Vida}
\affiliation{HUN-REN CSFK Konkoly Observatory, MTA Centre of Excellence, 
Konkoly Thege M. \'ut 15-17, Budapest, 1121, Hungary}

\begin{abstract}

We present new photometric observations of the core-collapse supernova SN 2023ixf occurred in M101, taken with the RC80 and BRC80 robotic telescopes in Hungary. The initial nickel mass from the late-phase bolometric light curve extending up to 400 days after explosion, is inferred as $M_{\rm Ni} = 0.046 \pm 0.007$ M$_\odot$. The comparison of the bolometric light curve with models from hydrodynamical simulations as well as semi-analytic radiative diffusion codes reveals a relatively low-mass ejecta of $M_{\rm ej} \lesssim 9$ M$_\odot$, contrary to SN~2017eaw, another H-rich core-collapse event, which had $M_{\rm ej} \gtrsim 15$ M$_\odot$. 
\end{abstract}


\section{Introduction} \label{sec:intro}

The discovery of SN~2023ixf \citep{itagaki23} at very early phase \citep[ $\lesssim 1$ day after first light,][]{hiramatsu23} in the nearby galaxy M101  \citep[$D_{\rm L} \simeq 6.85$ Mpc,][]{riess22} triggered a particularly intense study of its progenitor as well as the early-phase interaction between a H-rich ejecta of a core-collapse supernova with the confined circumstellar matter (CSM) envelope in the immediate vicinity of the exploding star. 
From pre-explosion space- and ground-based photometry, a red supergiant progenitor star ($\log(L/L_\odot) \sim 5$, $T_{\rm eff} \simeq 3500$--4000 K, $\log(R/R_\odot) \sim 2.7$) was identified, which was enshrouded by a thick, dusty CSM \citep{sorais23,kilpatrick23,jencson23,pledger23,niu23,qin23,neustadt24,xiang24,ransome24,vandyk24,zheng25,dickinson25,li25,shrestha25,jg25}. Interestingly, the initial mass ($M_{\rm ZAMS}$) of the progenitor turned out to be quite uncertain. It was estimated to be in a wide range of masses, from 9 to 20 M$_\odot$, and having a bimodal distribution: about half of the studies reported a relatively low $M_{\rm ZAMS}$ (11 -- 14 M$_\odot$), while others found it in between 16 -- 22 M$_\odot$ \citep[see Table~1 in][]{jg25}. The more massive progenitor ($M_{\rm prog} \sim 22$ M$_\odot$, implying an even higher $M_{\rm ZAMS}$) was also favored by \citet{liu23} using IFU spectroscopy and population synthesis of the surroundings of the SN site. This notable uncertainty in the progenitor mass is probably related to the presence of the thick, compact CSM that may have significantly affected the pre-explosion photometric data. 

Later, along with the accumulation of photometric/spectroscopic data of the SN itself, the estimation of the progenitor mass became possible via modeling the light curve (LC). As emphasized by \citet{bersten24}, LC-modeling could be a powerful tool that may be able to provide additional constraints on the physical parameters of the progenitor that are less affected by the uncertainties of the direct, pre-explosion measurements of the usually very faint progenitor. 

Most of the results from LC-modeling published so far favored the low-mass progenitor. \citet{hiramatsu23} found $M_{\rm prog} \sim 11$ M$_\odot$ implying $M_{\rm ZAMS} \sim 12$ M$_\odot$, similar to the results of \citet{bersten24} ($\sim 12$ M$_\odot$), \citet{singh24} ($\sim 10$ M$_\odot$), \citet{kozyreva25} ($\sim 10$ M$_\odot$), \citet{fang25} ($\lesssim 9.5$ M$_\odot$), \citet{cosentino25} ($\sim 9$ M$_\odot$). These studies compared the synthesized LCs of pre-existing \citep[e.g.][]{moriya24} or newly computed \citep{bersten24} hydrodynamical simulations with the observations, in order to constrain the ejecta (and the progenitor, via stellar evolution models) mass. The relatively low-mass progenitor is consistent with the results of \citet{zimmerman24}, who used the integrated bolometric luminosity of the SN to estimate the ejecta mass as $M_{\rm ej} \sim 7.5$ M$_\odot$, implying $M_{\rm ZAMS} \lesssim 10$ M$_\odot$. 

Some other studies, however, found somewhat higher progenitor masses. For example,
\citet{yang24} applied semi-analytic scaling relations to their $griz$ photometric data, and inferred a somewhat broader range for the possible ejecta mass, $M_{\rm ej} \sim 4$--16 M$_\odot$. 

On the other hand, based on new {\tt MESA+STELLA} model computations, \citet{hsu24} pointed out that the relatively low ejecta mass, $M_{\rm ej} \lesssim 8$ M$_\odot$, does not necessarily imply low ZAMS mass for the progenitor. Their best-fit models had initially $M_{\rm ZAMS} \sim 17.5$--21.5 M$_\odot$, but all of them lost almost half of their initial mass, and ended up having only $M_{\rm prog}=$7--8 M$_\odot$ at the moment of core collapse. They also proposed that it may be possible to break the degeneracy between the physical parameters inferred from photometry by detecting and measuring the pulsational variability of the progenitor prior to core collapse. Similarly, \citet{forde25} found that MESA+STELLA models having $M_{\rm prog} \sim 15$ M$_\odot$ and $M_{\rm ZAMS} \sim 23$ M$_\odot$ (implying significant mass loss) describe the LC at the end of the plateau phase better than lower mass progenitor models with less amount of mass loss.

After SN~2023ixf entered into the nebular phase, \citet{ferrari24} presented a new constraint on the initial mass of the progenitor based on nebular spectroscopy. From the line ratios of forbidden [OI] and [CaII] emission features, they inferred $12 < M_{\rm ZAMS} < 15$~M$_\odot$ for the initial mass of the progenitor. Note, however, that 12~M$_\odot$ was the lowest initial mass they considered, and the actual ejecta mass probed by the nebular spectra depends on the mass-loss history of the progenitor, which is uncertain. 
Since then many similar results were published, all within the range of $M_{\rm ZAMS} \sim 15 \pm 5$ M$_\odot$, by independent groups \citep{li25, fang25, zheng25, kumar25, michel25, folatelli25}.

The apparent diversity in the published ZAMS/progenitor masses of SN~2023ixf suggests that this issue is still far from being resolved.
In this paper we present an independent LC-modeling of SN~2023ixf by applying two different approaches: we use the semi-analytic code LC2\footnote{\url http://titan.physx.u-szeged.hu/$\sim$nagyandi/LC2.2/} \citep{nagy14, nagy16} and the hydrodynamical code SNEC\footnote{\url https://stellarcollapse.org/index.php/SNEC.html} \citep{morozova15} for synthesizing Type II SN LCs and comparing them with the bolometric data assembled from the observations. We also use SN~2017eaw, another H-rich Type II-P event, as a comparison object to highlight the similarities and differences between the physical parameters of the two SNe.

\section{Observations} \label{sec:obs}

\begin{figure}
    \centering
    \includegraphics[width=1.0\linewidth]{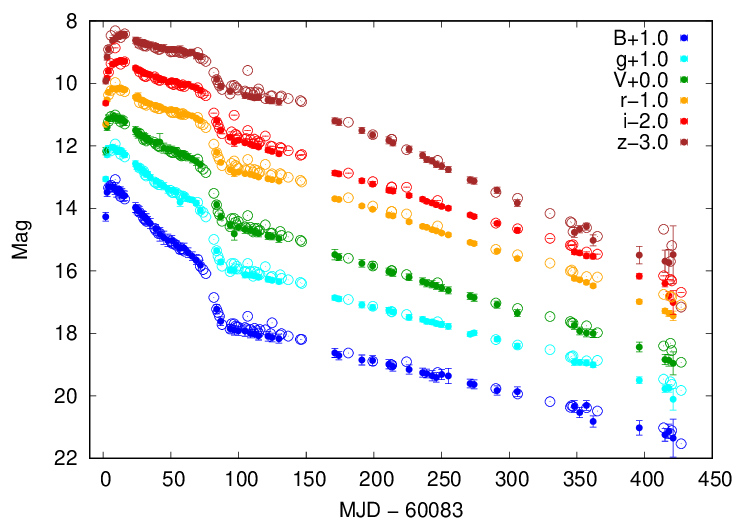}
    \caption{$BVgriz$ light curve of SN~2023ixf. Data from Konkoly and Baja are plotted with filled and open symbols, respectively. Different filters are color-coded, and a small vertical offset was applied between them for display purposes, as indicated in the legend. }
    \label{fig:lcobs}
\end{figure}

\begin{figure}
    \centering
    \includegraphics[width=1.0\linewidth]{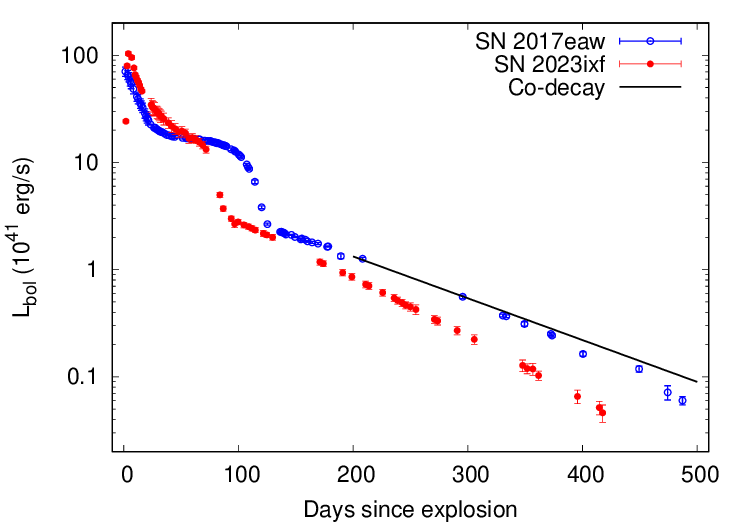}
    \caption{The pseudo-bolometric light curves of SN~2023ixf (red) and SN~2017eaw (blue). The black line represents the expected luminosity decline due to the radioactive decay rate of $^{56}$Co. }
    \label{fig:lc_comp}
\end{figure}

For SN~2023ixf, optical photometric data  were collected through $BVgriz$ filters with the two 0.8m robotic telescopes (RC80 and BRC80) operating at Piszkéstet{\H o} station of Konkoly Observatory and at Baja Observatory of University of Szeged, Hungary. For the description of the telescopes, as well as details on the data reduction and photometry, see \cite{barna23}. The $BVgriz$ light curve is shown in Figure~\ref{fig:lcobs}. It illustrates the previous finding that SN~2023ixf was a ``hybrid'' Type IIP/IIL event characterized by a shorter plateau and higher LC decline rate during the plateau phase \citep[e.g.][]{singh24}.

From the multi-band photometric data the spectral energy distribution (SED) was assembled for each epoch, and it was integrated over wavelength to get a pseudo-bolometric LC in a similar way as in \citet{szalai19}. For the ultraviolet (UV) fluxes we used {\it Swift} photometry collected from \citet{zimmerman24}. The UV fluxes had significant contribution on the integrated bolometric LC only during the first $\sim 30$ days after explosion. For the infrared (IR) part, a Rayleigh-Jeans tail was fitted to the $z$-band fluxes, and it was integrated between $\lambda_z$ and infinity to estimate the (unobserved) IR-contribution at each epoch. 

For the total (Milky Way and M101) interstellar extinction we applied $E(B-V) = 0.035$ mag from \citet{liu23}, which is based on high-resolution spectra of SN~2023ixf, and assumed standard reddening law with $R_V = 3.1$.

Finally, we adopted the luminosity distance of M101 from \cite{riess22} ($D_L \simeq 6.85$ Mpc), which is based on the most recent calibration of Cepheids, to get the luminosity as a function of time. {The uncertainty of the distance is estimated by \citet{vandyk24} as $\pm 0.32$ Mpc.}
The final pseudo-bolometric LC is plotted in Figure~\ref{fig:lc_comp}, together with that of SN~2017eaw taken from \cite{szalai19}. 

From Figure~\ref{fig:lc_comp} it is immediately seen that while the luminosity evolution of SN~2017eaw closely follows the expected decline due to the radioactive decay of $^{56}$Co (at least up to $\sim 450$ days) , it is not the case for SN~2023ixf. Such a faster luminosity decline is usually a consequence of a lower ejecta mass, when the SN envelope becomes progressively transparent to gamma-rays released by radioactive processes more quickly than in the case of a more massive ejecta. The physical parameters of these two SNe are modeled and compared in the following sections.

\section{Modeling}\label{sec:modeling}

This section contains the details of the semi-analytical and hydrodynamical models we built for SN~2023ixf. The results are presented in Section~\ref{results}.

\subsection{Nickel mass}
\label{sec:nimass}

For measuring the initial nickel mass, we fit the bolometric luminosity during the tail phase (i.e. after the end of the plateau at $t > 100$ days) with  
\begin{equation}
L(t) = \frac{M_{\rm Ni}}{M_\odot} \times \left ( L_\gamma(t) \times (1 - \exp[-\frac{T_\gamma}{t}]) + L_{+}(t) \right ),
\label{eq:lum}
\end{equation}
where 
\begin{equation}
L_\gamma(t) = C_1 \times e^{-t/8.8} + 0.968 \cdot C_2 \times e^{-t/111.3},  
\end{equation}
represents the energy input from the deposition of $\gamma$-rays produced by the Ni-Co radioactive decay, with $t$ being in days, $C_1 = 6.45 \times 10^{43}$ erg~s$^{-1}$ and  $C_2 = 1.45 \times 10^{43}$ erg~s$^{-1}$, while 
\begin{equation}
L_+(t) = 0.032 \cdot C_2 \times ( e^{-t/111.3} - e^{-t/8.8})
 \end{equation}
describes the contribution from the thermalization of the kinetic energy of positrons produced by the Co-decay \citep[e.g.][]{valenti08, bw17}. Our model has a very minor difference from that used by \citet{hsu24}, since we treat the contribution from positron trapping separately from the $\gamma$-rays.

The $\gamma$-ray deposition is parametrized by the exponent $T_\gamma$, which is connected with the ejecta mass and kinetic energy as 
\begin{equation}
T_\gamma = M_{\rm ej} \sqrt{\frac{9}{40 \pi} \frac{\kappa_\gamma}{E_{\rm kin}}}
\label{eq:tg}
\end{equation}
for a constant density ejecta \citep{bw17}. For the gamma-ray opacity we adopted $\kappa_\gamma = 0.028$ cm$^2$~g$^{-1}$ \citep{colgate80}. 

By fitting Eq.\ref{eq:lum} to the bolometric light curve of SN~2023ixf after $t>100$ days we get $M_{\rm Ni} = 0.046 \pm 0.007$ M$_\odot$ and $T_{\gamma} = 238 \pm 6$ days. The latter corresponds to a mass of $M_{\rm ej} = 7.3 \pm 0.2$ M$_\odot$ for a constant density ejecta and $E_{\rm kin} \sim 10^{51}$ erg via Eq.\ref{eq:tg}. For comparison, a similar analysis with the bolometric data of SN~2017eaw results in $M_{\rm Ni} = 0.054 \pm 0.006$ M$_\odot$, $T_\gamma \sim 557$ days and $M_{\rm ej} \sim 17.0$ M$_\odot$ (note that the formal errors for $T_{\gamma}$ are too high in this case, as the bolometric decline rate of SN~2017eaw is very close to the Co-decay rate that would correspond to $T_\gamma \rightarrow \infty$).  

Our inferred nickel mass is consistent with the results of \citet{michel25} ($M_{\rm Ni} = 0.049$ M$_\odot$), \citet{bersten24} ($0.05$ M$_\odot$), \citet{li25} ($0.059$ M$_\odot$), \citet{singh24} ($0.06$ M$_\odot$) and \citet{moriya24} ($\sim 0.04 - 0.06$ M$_\odot$). Other studies found slightly higher nickel masses, e.g. $M_{\rm Ni} = 0.07$ M$_\odot$ \citep{zimmerman24,hsu24,cosentino25} up to $M_{\rm Ni} = 0.098$ M$_\odot$ \citep{yang24}. Note that beside the photometric errors, the nickel mass estimates are directly influenced by the systematic uncertainty of the assumed distance. In this paper we adopted $D_L = 6.85 \pm 0.35$ Mpc (see Section~\ref{sec:obs}), thus, the relative error of the distance is $\delta D_L = 0.05$. This corresponds to a systematic uncertainty of the initial nickel mass of $\pm 0.002$ M$_\odot$, which is much lower than the scattering of the various nickel mass estimates appeared in the literature. Thus, it is probable that the source of the disagreement is mostly the uncertainties of the bolometric light curves used in different studies.

From this approximate analysis of the late-phase LC we conclude that, in accord with Fig.\ref{fig:lc_comp}, the faster luminosity decline of SN~2023ixf compared to that of SN~2017eaw can be explained by a factor of $\sim 2$ lower ejecta mass of SN~2023ixf, if the kinetic energy of the two events were similar to the canonical $E_{\rm kin} \sim 10^{51}$ erg. 

A more detailed modeling of the whole bolometric LC is presented below.

\subsection{Complete LC modeling}\label{full-lc-model}
\begin{figure}
    \centering
    \includegraphics[width=1.0\linewidth]{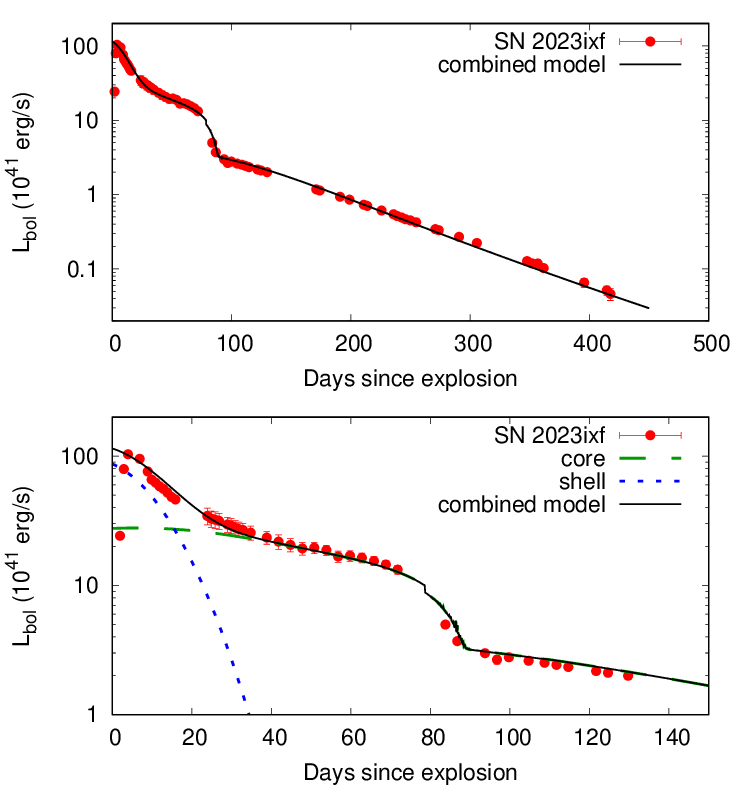}
    \caption{The bolometric LC of SN~2023ixf plotted together with the semi-analytic LC2 model shown in Table~\ref{tab:lc2}. The top panel shows the data with the combined (core+shell) model, while the bottom panel displays the core and shell components separately, as well. See text for explanation.}
    \label{fig:lc_23ixf}
\end{figure}

\begin{figure}
    \centering
    \includegraphics[width=1.0\linewidth]{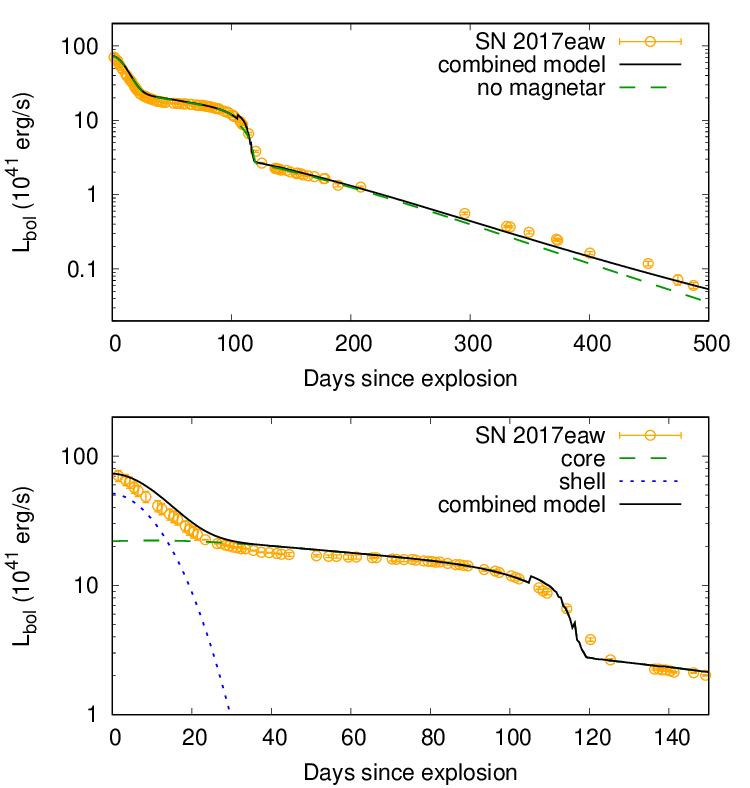}
    \caption{The same as Fig.~\ref{fig:lc_23ixf} but for SN~2017eaw. The dashed line in the upper panel represents the model without the energy input from the magnetar. }
    \label{fig:lc_17eaw}
\end{figure}

\begin{table*}
    \centering
    \caption{Parameters of the semi-analytic LC2 models}
    \begin{tabular}{lcc|cc}
    \hline \hline
     & SN~2023ixf & & SN~2017eaw & \\
    Parameter & shell & core & shell & core\\
    \hline
    Initial radius ($10^{13}$ cm) & 20 & 1.5 & 4.7 & 3.3 \\
    Ejecta mass ($M_\odot$) & 0.35 & 7.5 & 0.39 & 14.0 \\
    Nickel mass ($M_\odot$) & 0 & 0.046 & 0 & 0.043 \\
    Kinetic energy ($10^{51}$ erg) & 0.04 & 1.25 & 0.11 & 1.80 \\
    Thermal energy ($10^{51}$ erg) & 0.025 & 1.00 & 0.07 & 0.90 \\
    Recombination temperature (K) & 0 & 7500 & 0 & 7500 \\
    Mean opacity (cm$^2$~g$^{-1}$) & 0.34 & 0.2 & 0.34 & 0.2 \\
    Magnetar energy ($10^{51}$ erg) & 0 & 0 & 0 & 0.002 \\
    Magnetar spin-down (day) & 0 & 0 & 0 & 100 \\
    \hline
    \end{tabular}
    \label{tab:lc2}
\end{table*}

We utilized two different codes to model the entire bolometric light curve of SN~2023ixf. First, we used LC2 \citep{nagy14, nagy16}, which is a semi-analytic code based on the radiation-diffusion model of \citet{arnett80, arnett82, arnettfu89} including H-recombination. The latest version (LC2.2) was adopted, which, in addition to the radioactive decay of $^{56}$Ni--$^{56}$Co and H-recombination, contains an optional energy input from the spin-down of a magnetized neutron star \citep{kasen10}. We used the two-component ``core+shell'' model, where the ``shell'' component describes the first part ($\lesssim 30$ days) of the LC, while the ``core'' component models the rest of that. Since our model does not contain the energy released by the ejecta-CSM interaction that may significantly affect the early part of the LC, the ``shell'' parameters may not be physically self-consistent with those of the ``core'' component. 

Table~\ref{tab:lc2} lists the input parameters of LC2 as well as their best-fit values\footnote{All models were computed manually, without applying any automated fitting routine, due to the strong parameter degeneracy. Thus, the reported best-fit values represent only a possible solution for the strongly ill-constrained problem.} for SN~2023ixf and SN~2017eaw. The recombination temperature and the mean optical opacity for the core component were fixed at 7500 K and 0.2 cm$^2$~g$^{-1}$, respectively \citep[see][]{nagy16}. Since the low-mass shell component was assumed to be H-rich, its opacity was fixed at $\kappa = 0.34$ cm$^2$~g$^{-1}$, and H-recombination was ignored in the shell.  

Following the plateau phase, when the H-rich envelope recombines and becomes mostly transparent, the luminosity evolution is expected to follow the energy input due to the radioactive decay of $^{56}$Co. During this radioactive tail phase the observed luminosity decline is governed by the trapping of gamma-rays and positrons released by the Co-decay. For the gamma-ray opacity we adopted $\kappa_\gamma = 0.028$ cm$^2$~g$^{-1}$ \citep{colgate80}, while for positrons we applied an effective opacity of $\kappa_{+} = 7.0$ cm$^2$~g$^{-1}$ \citep{valenti08}. 

Secondly, we applied the 1-D Lagrangean hydrocode SNEC version 1.00 \citep{morozova15} to synthesize bolometric light curves from the simulated explosion of certain model stars. For the input models we used MESA version r-12778 \citep{paxton11, paxton13, paxton15, paxton18, paxton19, jermyn23} to construct models of $M_{\rm ZAMS} = 11, 12, 15, 20$ and 25 M$_\odot$ stars, and let them evolve until core collapse. 

We ran the 11 M$_\odot$ model (M11 hereafter), following the {\tt example\_make\_pre\_ccsn} test suite,  with $Z=0.006$ initial metallicity and used the "Dutch" hot wind-scheme with $\eta$ = 0.8 scaling factor \citep{nugis2000, gleb09}. For convection we considered overshooting by setting the diffusion mixing coefficient as $f_0 = 0.004$ and $f = 0.01$, and used the Ledoux criterion with 0.01 $\alpha$ semiconvection. For the 20 M$_\odot$ model (M20 hereafter) we followed
the {\tt 25M\_pre\_ms\_core\_collapse} test suite considering $Z=0.02$ initial metallicty and adopting the same wind scheme and convection as for the 11 M$_\odot$ model. In order to cover the parameter space for the LC fitting, we built further models with the same prescription as for M20. We created models with 12 M$_\odot$ (M12 hereafter), 15 M$_\odot$ (M15n hereafter) and 25 M$_\odot$ (M25 hereafter). Unlike the M11 model, M12 did not run into numerical difficulties when solar metallicity was assumed, thus, M12 was also built assuming $Z=0.02$. 
Note that the SNEC source code also contains a $M_{\rm ZAMS} = 15$ M$_\odot$ model (M15o hereafter) that was built with an earlier version of MESA. Since the main parameters of that model (progenitor mass and radius, density distribution, etc.) are different from those of our M15n model, we used M15o model as an input in SNEC as well.

In order to take into account the presence of the confined CSM, we attached an extended envelope on top of the MESA models using
the two-component CSM model given by \citet{kuri20},  \citet{tsuna21} and \citet{tsuna23}, which assumes a single eruption from the progenitor years before the explosion. We followed the interpretation of \citet{tsuna23} to construct the CSM as
\begin{equation}
{{\rho_{CSM}(r)} \over \rho_0} =  \left [ {1 \over 2} \left ( ({r \over r_{CSM}})^{1.5 \over y} + ({r \over r_{CSM}})^{n_{\rm out} \over y}\right ) \right ]^{-y}
\end{equation}
where $\rho_0 \approx M_{CSM} / (37 \cdot r_{CSM}^3)$, 
see Eq. (2) and Eq. (3) and their detailed description in \citet{tsuna23}. Based on \citet{tsuna21} we applied the parameters of an RSG progenitor, $y = 2$ and $n_{\rm out}$ = 12. This way we created a CSM model with a double power-law density profile, dividing the CSM to a bound and an unbound part, and attached it to the outmost part of the stellar envelope with no gap in between them. The key parameters of the CSM (total mass and outer radius) were set similar to the literature results cited above and also the mass and the radius of the ``shell'' component of the LC2 model.

In SNEC we adopted the ,,Thermal Bomb'' explosion model and the input physical parameters listed in Table~\ref{tab:snec}. The opacity floor of $\kappa_{\rm min}^{\rm core} = 0.24$ and $\kappa_{\rm min}^{\rm env} = 0.01$ cm$^2$~g$^{-1}$ was adopted for the core and the envelope, respectively.
The mass of the neutron star remnant ($M_{\rm NS}$) was assumed to be equal to the excised mass, an input parameter in SNEC. In some models it was set as 1.4 M$_\odot$ (the default value in SNEC), but for most of them it was defined to be equal to the mass of the iron core of the input MESA model.

The model parameters applied in the SNEC simulations are summarized in Table~\ref{tab:snec}. The ejecta masses are calculated as $M_{\rm ej} = M_{\rm prog}-M_{\rm NS}$. Since SNEC does not compute explosive nucleosynthesis, the initial mass of radioactive $^{56}$Ni is another input parameter. For that we adopted the nickel mass derived in Section~\ref{sec:nimass}, except for M20-3-csm and M25-csm, where $M_{\rm Ni} = 0.03$ M$_\odot$ was used in order to better match the early tail luminosity of the $M\gtrsim20$ M$_\odot$ models. In each model (except M15o-ni, see later) the distribution of $^{56}$Ni was assumed to extend up to 3 M$_\odot$ within the ejecta in mass coordinates. The corresponding maximum velocity ($v_{\rm Ni}^{\rm max}$) is given in Table~\ref{tab:snec} to indicate the degree of mixing in the ejecta.   The $E_{\rm tot}$ energy is the actual energy released by the thermal bomb, while $E_{\rm final}$ (also an input parameter) represents practically the asymptotic kinetic energy at late phases \citep{morozova18}. 

The last column of Table~\ref{tab:snec} indicates whether the boxcar smoothing algorithm were applied during the simulation: 1 means yes, 0 means no. Smoothing mimics the effect of mixing during the explosion, thus, it is generally applied in SNEC simulations \citep{morozova18}. The reason for switching it off in some models was the finding that it also makes the end of the plateau shallower and somewhat shorter, which is not consistent with the observed LC of SN~2023ixf \citep[see also][]{forde25}.

\section{Results and discussion}\label{results}

In this section we present and discuss the results from the two modeling approaches (semi-analytic vs. hydrodynamical) and their implications to the progenitor of SN~2023ixf.

\subsection{LC2 models}\label{results-lc2}

The output luminosities of the LC2 models that were found to best represent the observations are plotted in Fig.~\ref{fig:lc_23ixf} and \ref{fig:lc_17eaw} for SNe~2023ixf and 2017eaw, respectively. In these plots the top panel contains the whole observed light curve, while the bottom panel shows only the first 150 days, and the ``core'' and ``shell'' components alone are plotted as well. The ``shell'' component contributes only during the first 30 -- 40 days. 

As seen in Fig.~\ref{fig:lc_23ixf} and \ref{fig:lc_17eaw}, both the level and the length of the LC plateau are modeled very well by the LC2 code for both SN~2023ixf and SN~2017eaw. The same is true for the radioactive tail part, when the light curve decline rate is affected by the $^{56}$Co-decay as well as the gamma-ray leaking due to the dilution of the expanding ejecta. The key parameter for these two effects are the initial nickel mass ($M_{\rm Ni}$) and the ejecta mass ($M_{\rm ej}$) together with the kinetic energy ($E_{\rm kin}$) as detailed in Section~\ref{sec:nimass}. It is important to note that the latter two parameters also affect the LC during the plateau phase: higher $M_{\rm ej}$ makes the plateau longer, while higher $E_{\rm kin}$ shortens it \citep[e.g.][]{nagy14}. The fact that the LC2 model fits both the plateau and the tail simultaneously with the same set of parameters makes strong constraints on them, especially on the ejecta mass that we focus on. 

The constraint on the ejecta mass is also strengthened by the fit to the LC of SN~2017eaw. In that case both the plateau is longer and the decline during the tail is shallower, which is explained self-consistently by the larger $M_{\rm ej}$ (see Table~\ref{tab:lc2}). Note that the fitting of the late-phase decline needed a small amount of magnetar input in addition to the radioactive decay energy. Nevertheless, it is seen that SN~2017eaw very likely had a factor of $\sim 2$ higher ejecta mass than SN~2013ixf, which is consistent with the previous estimate in Section~\ref{sec:nimass}. 

It might be interesting that the initial radii of the outer ``shell'' components differ by almost an order of magnitude for the two SNe. For SN~2023ixf, the $R_0 \sim 2 \times 10^{14}$ cm value looks similar to the radius of the dense confined CSM \citep{zimmerman24, vandyk24}, even though it might be only a coincidence as the LC2 model assumes a constant density profile for the shell, while the density of confined CSM usually decreases outward. Nevertheless, the shell radius for SN~2017eaw turned out to be much smaller, which is consistent with the moderate early CSM interaction in that case.

\subsection{SNEC models}\label{results-snec}

\begin{figure*}
\centering
\includegraphics[width=0.9\linewidth]{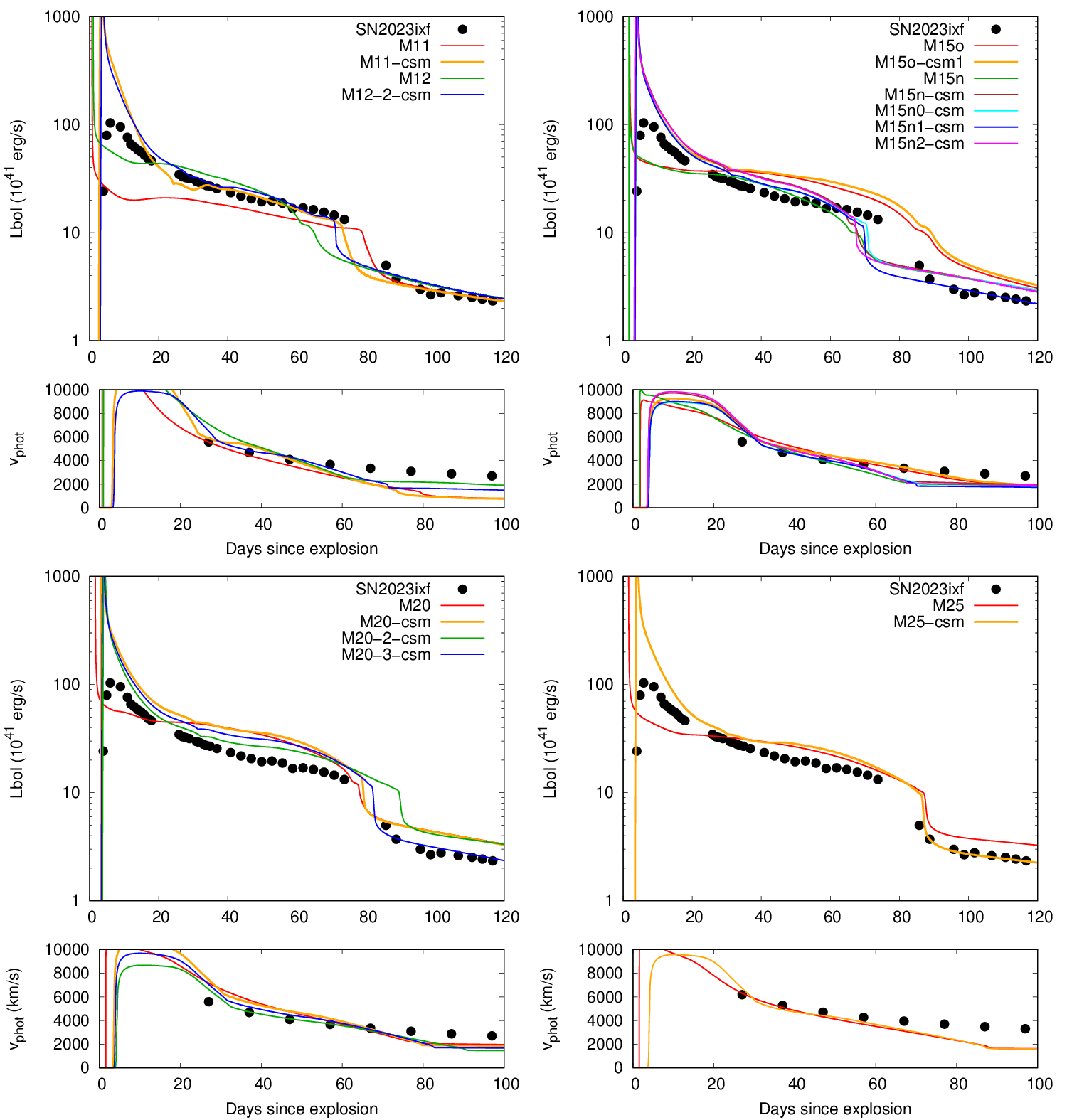}
\caption{SNEC model luminosities and photospheric velocities (colored curves) plotted together with the observed data (black circles). 
The observed velocities, derived from the FeII $\lambda 5189$ feature, were decreased by 600 km~s$^{-1}$ to account for the difference between the optical depths of the observed feature and the photosphere (see text). The model IDs are shown in the legends.}
\label{fig:snec}
\end{figure*}

\begin{table*}
\centering
\caption{Parameters of the SNEC models for SN~2023ixf. }
\begin{tabular}{lcccccccccccr}
\hline \hline
Model & $M_{\rm ini}$ & $M_{\rm prog}$ & $R_{\rm prog}$ & $M_{\rm csm}$ & $R_{\rm csm}$ & $M_{\rm Ni}$ & $M_{\rm NS}$ & $M_{\rm ej}$ & $E_{\rm tot}$ & $E_{\rm final}$ & $v_{\rm Ni}^{\rm max}$ & sm\\
& (M$_\odot$) & (M$_\odot$) & ($10^{13}$ cm) & (M$_\odot$) & ($10^{13}$ cm) & (M$_\odot$) & (M$_\odot$) & (M$_\odot$) & (foe) & (foe) & (km~s$^{-1}$) & \\
\hline
M11     & 11 & 10.26  & 3.04 &  --  &  --  & 0.046 & 1.40 & 8.86 & 2.23 & 1.20 & 1277 & 0 \\
M11-csm1 & 11 & 10.26 & 3.04 & 0.35 & 20.0 & 0.046 & 1.40 & 8.86 & 2.26 & 1.60 & 1448 & 0\\
M11-csm2 & 11 & 10.26 & 3.04 & 0.35 & 60.0 & 0.046 & 1.59 & 8.67 & 1.97 & 1.60 & 1392 & 1\\
M11-csm3 & 11 & 10.26 & 3.04 & 0.35 & 25.0 & 0.046 & 1.59 & 8.67 & 1.97 & 1.60 & 1392 & 1\\
M11-csm4 & 11 & 10.26 & 3.04 & 0.18 & 20.0 & 0.046 & 1.59 & 8.67 & 1.98 & 1.60 & 1393 & 1\\
M11-csm5 & 11 & 10.26 & 3.04 & 0.50 & 30.0 & 0.046 & 1.59 & 8.67 & 1.98 & 1.60 & 1399 & 1\\
\hline
M12     & 12 & 10.42  & 5.03 &  --  &  --  & 0.046 & 1.50 & 8.92 & 3.08 & 1.60 & 1985 & 1\\
M12-csm & 12 & 10.42  & 5.03 & 0.35 & 20.0 & 0.046 & 1.50 & 8.92 & 3.08 & 1.60 & 1977 & 1\\
M12-2-csm & 12 & 10.42 & 5.03 & 0.35 & 20.0 & 0.046 & 1.60 & 8.82 & 2.37 & 1.20 & 1737 & 0\\
\hline
M15o     & 15 & 12.29 & 7.23 &  --  &  --  & 0.046 & 1.40 & 10.89 & 1.84 & 1.20 & 1353 & 1\\
M15o-csm1 & 15 & 12.29 & 7.23 & 0.35 & 20.0 & 0.046 & 1.40 & 10.89 & 1.89 & 1.25 & 1384 & 1\\
M15o-csm2 & 15 & 12.29 & 7.23 & 0.35 & 60.0 & 0.046 & 1.60 & 10.69 & 1.58 & 1.20 & 1303 & 1\\
M15o-ni  & 15 & 12.29 & 7.23 & 0.35 & 20.0 & 0.046 & 1.40 & 10.89 & 1.89 & 1.25 & 5268 &1\\
M15n     & 15 & 11.69 & 6.60 & -- & --  & 0.046 & 1.50 & 10.19 & 2.18 & 1.20 & 1671 & 1\\
M15n-csm & 15 & 11.69 & 6.60 & 0.35 & 20 & 0.046 & 1.50 & 10.19 & 2.18 & 1.20 & 1853 & 1\\
M15n0-csm & 15 & 11.69 & 6.60 & 0.35 & 20 & 0.046 & 1.50 & 10.19 & 1.99 & 1.00 & 1695 & 0\\
M15n1-csm & 15 & 11.69 & 6.60 & 0.35 & 20 & 0.046 & 1.40 & 10.29 & 2.58 & 1.00 & 1639 & 0\\
M15n2-csm & 15 & 11.69 & 6.60 & 0.35 & 20 & 0.046 & 1.50 & 10.19 & 2.19 & 1.20 & 1876 & 0\\
\hline
M20     & 20 & 15.85 & 6.83  &  --  &  --  & 0.046 & 1.40 & 14.45 & 4.30 & 1.80 & 1672 & 1\\
M20-csm & 20 & 15.85 & 6.83 & 0.35 & 20.0 & 0.046 & 1.40 & 14.45 & 4.30 & 1.80 &  1699 & 0\\
M20-2-csm & 20 & 15.85 & 6.83 & 0.35 & 20.0 & 0.046 & 1.76 & 14.09 & 2.73 & 1.20 & 1305 & 0\\
M20-3-csm & 20 & 15.85 & 6.83 & 0.35 & 20.0 & 0.030 & 1.76 & 14.09 & 3.03 & 1.50 & 1527 & 0\\
M20-csm2 & 20 & 15.85 & 6.83 & 0.35 & 60.0 & 0.046 & 1.76 & 14.09 & 3.32 & 1.80 & 1672 & 1\\
\hline
M25     & 25 & 20.75 & 5.78  &  --  &  --  & 0.046 & 1.99 & 18.76 & 3.05 & 1.80 & 1391 & 0\\
M25-csm & 25 & 20.75 & 5.78 & 0.35 & 20.0 & 0.030 & 1.99 & 18.76 & 3.05 & 1.80 &  1363 & 0\\
\hline
\end{tabular}
\tablecomments{The last column indicates whether {\tt boxcar{\_}smoothing} was applied (1) or not (0) in the model.}
\label{tab:snec}
\end{table*}

Figure~\ref{fig:snec} presents the evolution of the luminosities (top panels) and the photospheric velocities (bottom panels) for the M11/M12, M15, M20 and M25 SNEC models, respectively, together with the observations for SN~2023ixf. For the luminosities we used the pseudo-bolometric LC derived in Section~\ref{sec:obs}, while for the velocities we adopted the fit to the observed velocities from \citet{singh24} as $v_{\rm obs} \simeq 4350 \cdot (t_d / 53)^{-0.47}$, where $t_d$ is the time since explosion in days, for $t_d > 20$ days.  

It is known that the observed velocities, usually calculated from the Doppler-shift of the absorption minimum of certain features, like FeII $\lambda5169$ or $H\beta$, usually overestimate the photospheric velocities inferred directly from hydrodynamical models assuming $\tau_{\rm ph}=1$ or 2/3 as the optical depth of the photosphere \citep[e.g.][]{goldberg19,barker22}. In order to take this effect into account, the observed velocities were corrected as $v_{\rm phot} \approx v_{\rm obs} - 600$~km~s$^{-1}$ before plotting in Figure~\ref{fig:snec}.

It is seen that all models containing the extended CSM have the luminosity peak at least an order of magnitude brighter than the observations \citep[see also][]{kozyreva25}. Thus, in the followings we consider only the $t > 30$ days phases when comparing the model LCs with the data \citep{forde25}. We do not attempt to optimize the parameters of the CSM to fit the early-phase data, because $i)$ this topic has been already explored extensively in the literature, and $ii)$ SNEC may not be the best modeling code to predict the luminosity of the shock wave propagating in such environments, where, for example, the deviation from LTE may be substantial. Even though we focus on modeling the LC after shock breakout from the confined CSM, we explore the effect of varying the CSM parameters on the behavior of the late-phase LC in the next subsection.    

Figure~\ref{fig:snec} shows that the M15o, M20 and M25 models, both with and without the CSM envelope, overpredict the luminosity for the plateau phase after $t \gtrsim 30$ days. The M15n models are closer to the observed plateau luminosity, but they start the transition to the tail phase about $\sim 10$ days earlier than observed. 

Moreover, these high-mass models, all having $M_{\rm ej} > 10$ M$_\odot$, are also not consistent with the luminosity during the early tail phase: for $t > 80$ days all of them, except M15n1-csm, M20-3-csm and M25-csm, predict slightly higher luminosities than observed. At first this may be surprising, because all these models, except M20-3-csm and M25-csm, were computed with the same amount of $M_{\rm Ni}$ (0.046 M$_\odot$) that represents the best-fit to the observed tail light curve presented in Section~\ref{sec:nimass}. The probable reason for this discrepancy is that in these models the luminosity at the beginning of the tail phase (between $100 < t < 120$ days) has not yet settled to the level of the radioactive decay. Thus, the model LC shortly after the end of the plateau may still contain some energy input from the remaining photosphere. This means that the model LC at these early tail phases may underestimate the Ni-masses when compared to the observations. This is illustrated by the models M20-3-csm and M25-csm that produce better fits to the data at the early tail phases: these models were computed assuming $M_{\rm Ni} = 0.03$ M$_\odot$, instead of 0.046 M$_\odot$ used in all the other models.

The M15n1-csm model is unique in a sense that it fits the early tail luminosity even though it was computed with $M_{\rm Ni} = 0.046$ M$_\odot$, thus, it seems to contradict with the above statement. However, in this model the excised mass (practically the mass of the remnant neutron star) was set at 1.4 M$_\odot$, instead of 1.5 M$_\odot$, the latter being the mass of the iron core in the M15n model family. The output of this model illustrates some model-dependency of the tail luminosity predicted by SNEC: it is sensitive to the value of the parameter {\tt mass\_excised}. Exploring this effect more thoroughly is, however, beyond the scope of this paper.

The LCs from the M11 and M12 model families (having $M_{\rm ej} < 9$ M$_\odot$) are the most similar to the observed bolometric LC at the late plateau phase, after $\sim 30$ days, as well as during the radioactive tail phase (Figure~\ref{fig:snec} top left panel). The M11-csm model looks like the optimal one, because it fits both the mid-plateau luminosity as well as the length of the plateau adequately. 

Regarding the velocities shown in Figure~\ref{fig:snec}, all of the SNEC models considered here are more-or-less consistent with the observed velocities, at least for $t < 60$ days, if one takes into account the well-known offset between the observed and the model velocities ($\sim 600$ km~s$^{-1}$, see above). This is consistent with the result of \citet{goldberg19} who found that ``ejecta velocities measured during the majority of the plateau phase provide little additional information about explosion characteristics''. On the other hand, in principle, models that predict velocities that are clearly incompatible with the observed ones can be ruled out \citep[see e.g.][for an example]{uc24}. None of the models considered here, however, fall into this category based on the velocity curve alone.

\subsubsection{Comparison with STELLA models}\label{sec:stella}

\begin{table}[]
    \centering
    \caption{Observable parameters inferred from SNEC models (see text). The last row gives the actually observed values from the bolometric LC and velocity curve of SN~2023ixf.}
    \begin{tabular}{lccc}
    \hline
    \hline
       Model  &  $t_p$ & $L_{50}$ & $v_{50}$ \\
              &  (days) & ($10^{42}$ erg~s$^{-1}$) & (km~s$^{-1}$) \\
       \hline       
       M11    &   81 & 1.50 & 3298 \\
       M11-csm1 & 74 & 2.06 & 3752 \\
       M11-csm2 & 73 & 2.27 & 3927 \\
       M11-csm3 & 72 & 2.16 & 3752 \\
       M11-csm4 & 72 & 2.02 & 3637 \\
       M11-csm5 & 73 & 2.42 & 3952 \\
       \hline
       M12 & 62 & 2.40 & 3973 \\
       M12-csm & 63 & 2.66 & 4249 \\
       M12-2-csm & 71 & 2.27 & 4023 \\
       \hline
       M15o &  85 & 3.26 & 4336 \\
       M15o-csm1 & 87 & 3.32 & 4345 \\
       M15o-csm2 & 86 & 3.74 & 4771 \\
       M15o-ni & 84 & 3.45 & 4369 \\
       M15n & 66 &  2.15 & 3701 \\
       M15n-csm & 66 & 2.75 & 4124 \\
       M15n0-csm & 70 & 2.37 & 3894 \\
       M15n1-csm & 70 & 2.42 & 3922 \\
       M15n2-csm & 67 & 2.69 & 4115 \\
       \hline
       M20 & 77 & 3.43 & 4492 \\
       M20-csm & 79 & 3.58 &  4612 \\
       M20-2-csm & 89 & 2.66 & 4023 \\
       M20-3-csm & 82 & 3.12 & 4345 \\
       M20-csm2 & 79 & 4.20 & 4989 \\
       \hline
       M25 & 87 & 2.66 & 4134 \\
       M25-csm & 87 & 2.80 & 4299 \\
       \hline 
       SN2023ixf & 81 & 1.96 & 3750 \\
       \hline
    \end{tabular}
    \label{tab:snec_obs}
\end{table}

\begin{figure*}
    \centering
    \includegraphics[width=0.8\linewidth]{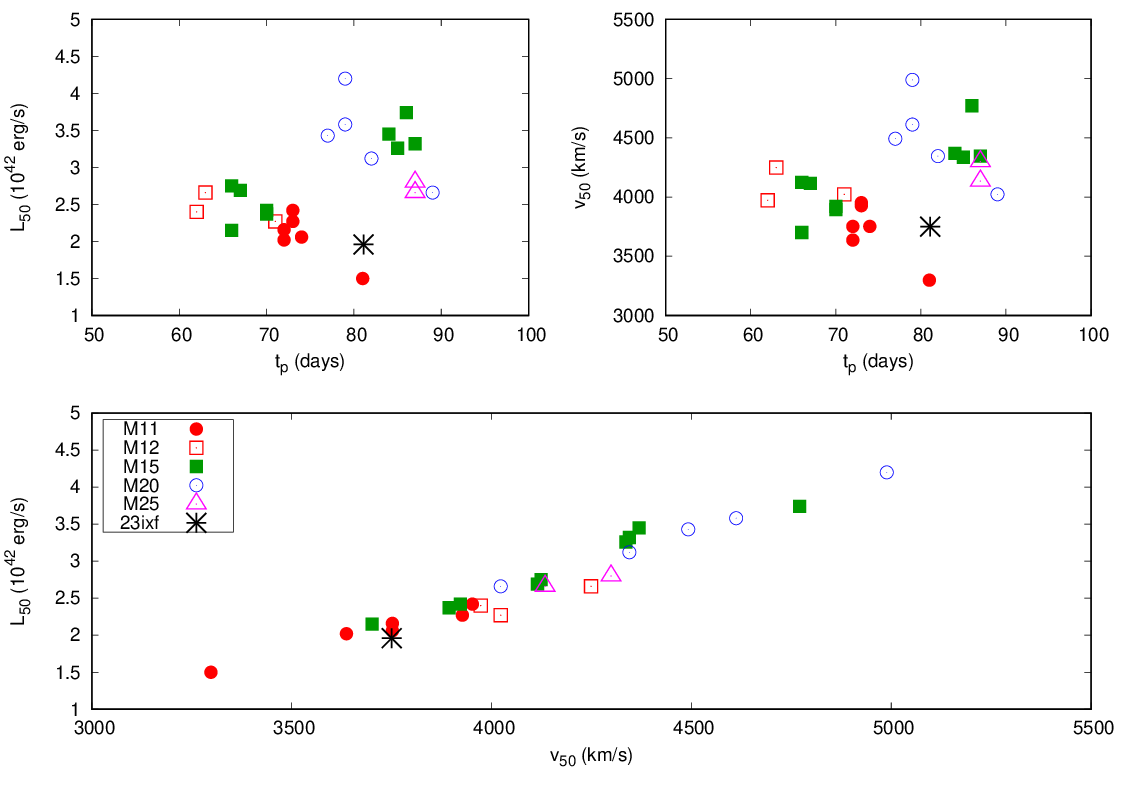}
    \caption{The observable parameters from SNEC models (see Table~\ref{tab:snec_obs}) compared with the observed values of SN~2023ixf (asterisk).}
    \label{fig:snec_obs}
\end{figure*}

\begin{figure*}
    \centering
    \includegraphics[width=0.8\linewidth]{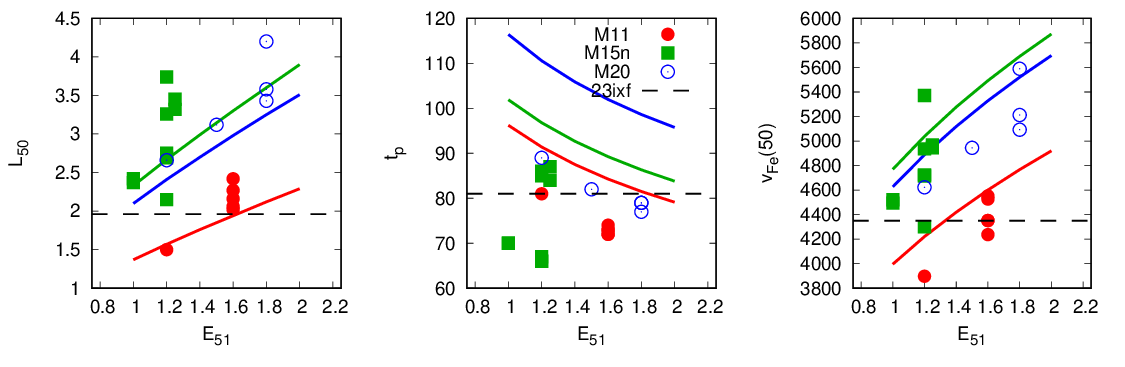}
    \caption{Comparison of the observables from SNEC (symbols) and from STELLA (curves) against the kinetic energy $E_{51}$. The dashed lines indicate the observed values for SN~2023ixf.}
    \label{fig:stella_obs}
\end{figure*}

In order to make more quantitative statements about the similarity between the observed luminosity/velocity curves and those from the models, first, we measured the key observable parameters of the SNEC LCs in order to compare them with the actual observed ones. Here the following parameters are considered, adopting the definitions by \citet{goldberg19}): the length of the plateau ($t_p$) in days; the luminosity at day +50 ($L_{50}$) in $10^{42}$ erg~s$^{-1}$ and the velocity at the photosphere at day +50 ($v_{50}$) in km~s$^{-1}$. These parameters, inferred from the SNEC models, are collected in Table~\ref{tab:snec_obs} and plotted in Figure~\ref{fig:snec_obs}. In the last row we show the actually observed values for SN~2023ixf (plotted with an asterisk), where the measured velocity has been corrected by $-600$ km~s$^{-1}$ as explained above.

Second, we applied the analytical formulae of \citet{goldberg19} to estimate the observable parameters from STELLA \citep{blinnikov98} models. The following parameters, taken from the input progenitor models for SNEC, are used in these calculations: the radius of the progenitor in 500 R$_\odot$ units ($R_{500}$), the ejecta mass in 10 M$_\odot$ units ($M_{10}$), the nickel mass in M$_\odot$ ($M_{\rm Ni}$) and the final energy of the explosion in $10^{51}$ erg ($E_{51}$). We considered $E_{51}$ as a free parameter within the range of $1 < E_{51} < 2$ covered by our SNEC models, and estimated $t_p$, $L_{50}$ and $v_{\rm Fe} \approx v_{\rm phot} + 600$ km~s$^{-1}$ using the equations given by \citet{goldberg19}. The results are plotted together with the SNEC observables in Figure~\ref{fig:stella_obs}.

From Figure~\ref{fig:snec_obs} it is seen that the observable parameters inferred from our SNEC models suggest the same conclusion as the visual inspection of the LCs and the velocities did in the previous section: the parameters of SN~2023ixf are most similar to the observables of the M11 models, even though no model gives a perfect match. The high-mass models (M15o, M20 and M25) predict too high $L_{50}$ and $v_{50}$ values. The M15n models have similar $L_{50}$ than the M11 and M12 models, but their $t_p$ values are too low compared to the observed plateau length. It is also clear from the velocity-luminosity relation (bottom panel in Figure~\ref{fig:snec_obs}) that SN~2023ixf is most consistent with the predictions of the low-mass SNEC models. 

Very similar conclusion can be drawn from the comparison with STELLA models in Figure~\ref{fig:stella_obs}. Even though STELLA generally predicts longer plateau phases than SNEC (as shown in the middle panel), it is seen that the observed parameters of SN~2023ixf, indicated by the horizontal dashed line in each panel, are closer to the red curves that correspond to the progenitor parameters of the M11 models. The agreement is best within the range of $1.2 < E_{51} < 1.6$, which is also consistent with the results from SNEC models.  

It is concluded that the comparison of the observable parameters predicted by both SNEC and STELLA hydrodynamical models suggests that SN~2023ixf had a relatively low-mass ejecta, $M_{\rm ej} < 9$ M$_\odot$ after the core collapse of an $M \sim$10-11 M$_\odot$ progenitor.

\subsection{Effects of different CSM parameters on the bolometric LC}

\begin{figure}
    \centering
    \includegraphics[width=1.0\linewidth]{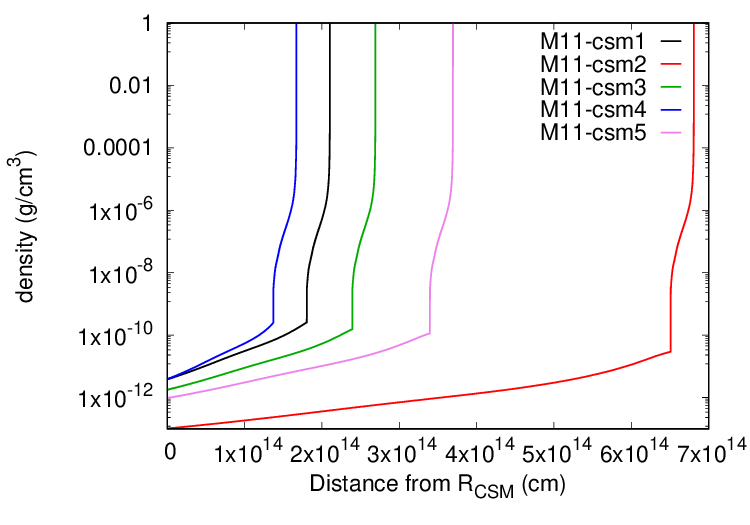}
    \includegraphics[width=1.0\linewidth]{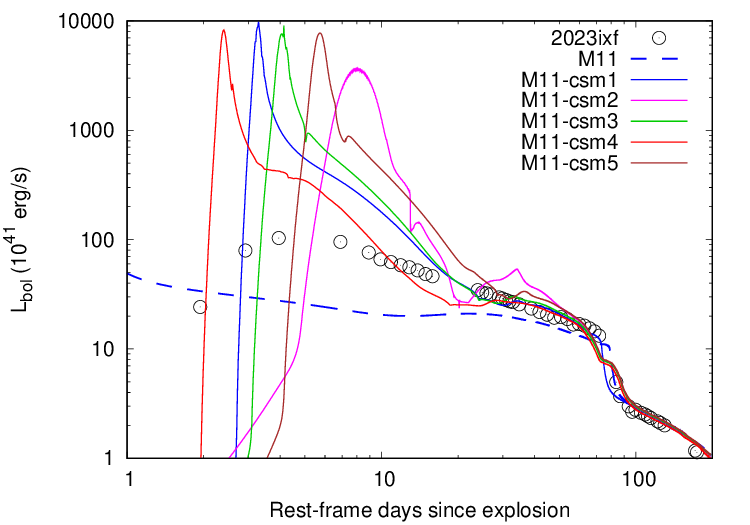}
    \caption{Top panel: The density distribution of the various M11-csm models listed in Table~\ref{tab:snec}. Note that the distance from the outmost surface of the CSM envelope (in cm) is plotted on the horizontal axis due to the convention used by SNEC. The innermost part of the SN envelope is omitted for display purposes. Bottom panel: The luminosity evolution of the various M11-csm models (colored curves) compared with the observations (red symbols). The LC from the M11 model without the CSM (dotted curve) is also shown for comparison. }
    \label{fig:snec-csm}
\end{figure}

Since SN~2023ixf was embedded in a dense, confined CSM that significantly affected the early LC as well as the spectral evolution during the first 10-15 days after explosion, we attempted to investigate whether adding such a CSM on top of the progenitor model in SNEC has an overall impact on estimating the ejecta physical parameters from LC modeling.  Details on the attached CSM are given in Section~\ref{full-lc-model}.  Following \citet{morozova16}, the mass and outer radius of the attached CSM ($M_{\rm csm}$ and $R_{\rm csm}$, respectively) were varied in between 0.35 -- 0.50 $M_\odot$ and 20 -- $60 \times 10^{13}$ cm, respectively (see Table~\ref{tab:snec}). The choice of these parameter ranges were motivated by the results of our semi-analytic modeling, namely the parameters of the ``shell'' component, given in Table~\ref{tab:lc2}. The density profiles of the attached CSM models are shown in the top panel of Figure~\ref{fig:snec-csm}. 

The final LCs of the M11 model family (see Table~\ref{tab:snec}) are plotted in the bottom panel of Figure~\ref{fig:snec-csm}, together with the observations. It is apparent that the early peaks of the model LCs are too high, by a factor of $\sim 100$, compared to the observations. Note that the sharp first peak of the SNEC model LCs corresponds to very high temperatures shortly after the shock breakout, which might be visible only in X-rays, thus, they are missing from the calculated bolometric luminosity of SN~2023ixf. But, even if we disregard the first sharp peak, all SNEC model LCs that contain CSM are still too bright during the first $\sim 20$ days after explosion. All the other models containing CSM show similar behavior (see Figure~\ref{fig:snec}). Thus, it seems that the assumed mass and radius of the CSM, similar to those of the ``shell'' component in Table~\ref{tab:lc2}, gives a CSM that is too dense to be compatible with the observed LC.

On the other hand, Figure~\ref{fig:snec-csm} also shows that after $\sim 30$ days the LCs containing the CSM contribution start to become similar to the LC without the CSM, in agreement with the calculations presented by \citet{morozova16}. Thus, instead of optimizing the CSM parameters to better fit the early part of the LC, we focus on the later part of the plateau, after $\sim 30$ days, when the contribution from the confined CSM becomes weak. This is also the phase when the first P Cygni features of H and He started to appear in the optical spectrum \citep[e.g.][]{teja23, bostroem24}, suggesting the emergence of the SN ejecta from behind the CSM.  
 
These calculations suggest that the ejecta mass estimate based on the bolometric LC after $\sim 30$ days is not sensitive to the presence or absence of a dense, confined CSM, since the CSM contribution on the LC becomes weak after this phase. Therefore, our conclusion that the M11-csm1 model gives the most similar LC to the observations of SN~2023ixf (Sections~\ref{results-snec} and \ref{sec:stella}) becomes unaffected. 

\subsection{Effects of extended $^{56}$Ni distribution on the bolometric LC}

\begin{figure}
    \centering
    \includegraphics[width=1.0\linewidth]{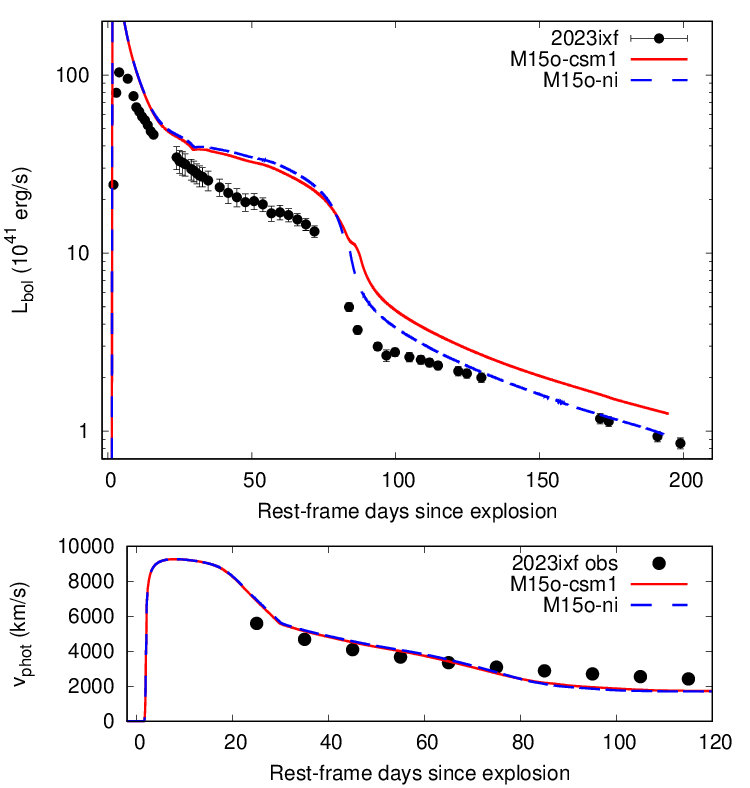}
    \caption{The effect of the extended Ni-distribution (model M15o-ni, plotted with blue dashed line) on the light curve (top panel) and on the velocity curve (bottom panel) compared with the observations (black circles) and the M15o-csm1 model (red lines). }
    \label{fig:snec-extni}
\end{figure}

Even though the length of the LC plateau ($t_p$) is found to be a useful indicator for the ejecta mass \citep[e.g.][]{litvinova85, nadyozhin03, nagy14}, the relation between them is not unique: 
$\Delta t_{\rm P}$ also depends on the explosion energy, the progenitor radius and other parameters as well \citep[see e.g.][]{nagy14,goldberg19,utrobin21}). Recently, a particularly interesting case was explored by \citet{uc24}: from hydrodynamical modeling they found that the short plateau length of SN~2018gj ($t_p \sim 75$ days) can be explained by a relatively massive ejecta, $M_{\rm ej} \gtrsim 20$~M$_\odot$, and an asymmetric distribution of $^{56}$Ni extending to high velocities ($v_{\rm Ni}^{\rm max} \gtrsim 5000$ kms$^{-1}$).  Similar bipolar Ni-distribution was found also in SN~2017gmr \citep{utrobin21}. In such a bipolar Ni-distribution some of the $\gamma$-rays released by the radioactive decay closer to the photosphere may escape more easily than the photons created deeper within the ejecta. Thus, $\gamma$-ray leaking becomes stronger, even during the plateau phase, resulting in less efficient heating of the ejecta, which enhances hydrogen recombination and reduces the plateau length. 

Since the plateau length was one of the observables that was used to constrain the ejecta mass of SN~2023ixf in Sect. \ref{results-lc2} and \ref{results-snec}, we attempted to address this issue with SNEC, even though the applicability of this 1D-hydro code for modeling this problem is very limited. Nevertheless, we computed a variant of the M15o-csm1 model (see Table~\ref{tab:snec}) with the outer mass coordinate of the Ni-distribution extending up to 12~M$_\odot$, i.e. close to the surface of exploding star (model M15o-ni). The resulting LC is plotted in Figure~\ref{fig:snec-extni} together with the observations and the original M15o-csm1 LC. 

It is seen that even within this simple model the extended Ni-distribution reduces the length of the plateau as well as the LC level during the early tail phase, but its effect is not strong enough: the shortening of $t_p$ is only a few days compared to the plateau length of the original model having centrally located $^{56}$Ni. Thus, the resulting LC is still less compatible with the observations of SN~2023ixf than the LCs produced by the models having less massive ejecta. It is concluded that while the Ni-distribution can certainly affect the fundamental properties of the model LCs, this effect is less pronounced in SNEC models, probably due to its simplified 1D (spherical) geometry. A more comprehensive study of this problem, which requires at least 2D modeling, is beyond the scope of the present paper.  

\section{Conclusions}

By modeling the bolometric LC constructed from observations, we confirm the relatively low-mass ejecta of SN~2023ixf ($M_{\rm ej} \sim 8$ -- 9~M$_\odot$), consistent with the results of \citet{hiramatsu23}, \citet{bersten24}, \citet{singh24}, \citet{hsu24} and \citet{forde25}.

Applying the semi-analytic LC2 code we get $\sim 7.8$ M$_\odot$ for the ejecta (core + shell) mass, with an uncertainty range of $\sim 1$ M$_\odot$, which, together with the other physical parameters listed in Table~\ref{tab:lc2}, is able to explain both the plateau length as well as the late-phase decline rate self-consistently. A comparison with SN~2017eaw, which showed both a longer plateau and a shallower late-phase decline, revealed that SN~2013ixf had a factor of $\sim 2$ less massive ejecta than SN~2017eaw. Taking into account the mass of the neutron star ($\sim 1.5$~M$_\odot$) as well as some mass-loss during the evolution after the main sequence, these results are consistent with a progenitor mass of $M_{\rm prog} \sim 10$~M$_\odot$ and $\sim17$~M$_\odot$ for SN~2023ixf, and SN~2017eaw, respectively. 

The synthesized LCs from SNEC simulations of massive stars fully support these results (see Table~\ref{tab:snec}). Even though the SNEC LCs do not extend to sufficiently late phases, the luminosity and the length of the plateau of the simulated LC constrains the ejecta mass. From these SNEC models the most likely ejecta mass, $\sim 9$~M$_\odot$, is somewhat higher but similar to the value provided by the LC2 code. 

\begin{acknowledgments}

The operation of the RC80 and BRC80 telescopes were supported by the GINOP 2.3.2-15-2016-00033 project of the National Research, Development and Innovation Office (NKFIH), Hungary, based on funding from the European Union. 
This research was supported by NKFIH OTKA grants K-142534, K-138962, FK-134432 and NKFIH KKP-143986 grant.
L.K. acknowledges the support of the NKFIH grant PD-134784. K.V. is supported by the Bolyai János Research Scholarship of the Hungarian Academy of Sciences.
A.B., C.K. and A.S were supported by the `SeismoLab' KKP-137523 \'Elvonal grant of the NKFIH. Zs.B. was supported by the ÚNKP-22-2 New National Excellence Program of the Hungarian Ministry for Culture and Innovation from the source of the NKFIH Fund.
N.O.SZ. thanks the financial support provided by the undergraduate research assistant program of Konkoly Observatory. 

\end{acknowledgments}

\vspace{5mm}
\facilities{RC80, BRC80}

\software{LC2 \citep{nagy16};
          SNEC \citep{morozova15}
          }

\bibliography{ms_rev2}{}

\begin{thebibliography}{}
\expandafter\ifx\csname natexlab\endcsname\relax\def\natexlab#1{#1}\fi
\providecommand{\url}[1]{\href{#1}{#1}}
\providecommand{\dodoi}[1]{doi:~\href{http://doi.org/#1}{\nolinkurl{#1}}}
\providecommand{\doeprint}[1]{\href{http://ascl.net/#1}{\nolinkurl{http://ascl.net/#1}}}
\providecommand{\doarXiv}[1]{\href{https://arxiv.org/abs/#1}{\nolinkurl{https://arxiv.org/abs/#1}}}

\bibitem[{{Arnett}(1980)}]{arnett80}
{Arnett}, W.~D. 1980, \apj, 237, 541, \dodoi{10.1086/157898}

\bibitem[{{Arnett}(1982)}]{arnett82}
---. 1982, \apj, 253, 785, \dodoi{10.1086/159681}

\bibitem[{{Arnett} \& {Fu}(1989)}]{arnettfu89}
{Arnett}, W.~D., \& {Fu}, A. 1989, \apj, 340, 396, \dodoi{10.1086/167402}

\bibitem[{{Barker} {et~al.}(2022){Barker}, {Harris}, {Warren}, {O'Connor}, \&
  {Couch}}]{barker22}
{Barker}, B.~L., {Harris}, C.~E., {Warren}, M.~L., {O'Connor}, E.~P., \&
  {Couch}, S.~M. 2022, \apj, 934, 67, \dodoi{10.3847/1538-4357/ac77f3}

\bibitem[{{Barna} {et~al.}(2023){Barna}, {Nagy}, {Bora}, {Czavalinga},
  {K{\"o}nyves-T{\'o}th}, {Szalai}, {Sz{\'e}kely}, {Zs{\'\i}ros},
  {B{\'a}nhidi}, {B{\'\i}r{\'o}}, {Cs{\'a}nyi}, {Kriskovics}, {P{\'a}l},
  {Szab{\'o}}, {Szak{\'a}ts}, {Vida}, {Bodola}, \& {Vink{\'o}}}]{barna23}
{Barna}, B., {Nagy}, A.~P., {Bora}, Z., {et~al.} 2023, \aap, 677, A183,
  \dodoi{10.1051/0004-6361/202346395}

\bibitem[{{Bersten} {et~al.}(2024){Bersten}, {Orellana}, {Folatelli},
  {Martinez}, {Piccirilli}, {Regna}, {Rom{\'a}n Aguilar}, \&
  {Ertini}}]{bersten24}
{Bersten}, M.~C., {Orellana}, M., {Folatelli}, G., {et~al.} 2024, \aap, 681,
  L18, \dodoi{10.1051/0004-6361/202348183}

\bibitem[{{Blinnikov} {et~al.}(1998){Blinnikov}, {Eastman}, {Bartunov},
  {Popolitov}, \& {Woosley}}]{blinnikov98}
{Blinnikov}, S.~I., {Eastman}, R., {Bartunov}, O.~S., {Popolitov}, V.~A., \&
  {Woosley}, S.~E. 1998, \apj, 496, 454, \dodoi{10.1086/305375}

\bibitem[{{Bostroem} {et~al.}(2024){Bostroem}, {Sand}, {Dessart}, {Smith},
  {Jha}, {Valenti}, {Andrews}, {Dong}, {Filippenko}, {Gomez}, {Hiramatsu},
  {Hoang}, {Hosseinzadeh}, {Howell}, {Jencson}, {Lundquist}, {McCully},
  {Mehta}, {Meza-Retamal}, {Pearson}, {Ravi}, {Shrestha}, \&
  {Wyatt}}]{bostroem24}
{Bostroem}, K.~A., {Sand}, D.~J., {Dessart}, L., {et~al.} 2024, \apjl, 973,
  L47, \dodoi{10.3847/2041-8213/ad7855}

\bibitem[{{Branch} \& {Wheeler}(2017)}]{bw17}
{Branch}, D., \& {Wheeler}, J.~C. 2017, {Supernova Explosions},
  \dodoi{10.1007/978-3-662-55054-0}

\bibitem[{{Colgate} {et~al.}(1980){Colgate}, {Petschek}, \&
  {Kriese}}]{colgate80}
{Colgate}, S.~A., {Petschek}, A.~G., \& {Kriese}, J.~T. 1980, \apjl, 237, L81,
  \dodoi{10.1086/183239}

\bibitem[{{Cosentino} {et~al.}(2025){Cosentino}, {Pumo}, \&
  {Cherubini}}]{cosentino25}
{Cosentino}, S.~P., {Pumo}, M.~L., \& {Cherubini}, S. 2025, \mnras, 540, 2894,
  \dodoi{10.1093/mnras/staf861}

\bibitem[{{Dickinson} {et~al.}(2025){Dickinson}, {Milisavljevic}, {Garretson},
  {Dessart}, {Margutti}, {Chornock}, {Subrayan}, {Hillier}, {Golub}, {Li},
  {Logsdon}, {Rajagopal}, {Ridgway}, {Smith}, \& {Cynamon}}]{dickinson25}
{Dickinson}, D., {Milisavljevic}, D., {Garretson}, B., {et~al.} 2025, \apj,
  984, 71, \dodoi{10.3847/1538-4357/adc108}

\bibitem[{{Fang} {et~al.}(2025){Fang}, {Moriya}, {Ferrari}, {Maeda},
  {Folatelli}, {Ertini}, {Kuncarayakti}, {Andrews}, \& {Matsumoto}}]{fang25}
{Fang}, Q., {Moriya}, T.~J., {Ferrari}, L., {et~al.} 2025, \apj, 978, 36,
  \dodoi{10.3847/1538-4357/ad8d5a}

\bibitem[{{Ferrari} {et~al.}(2024){Ferrari}, {Folatelli}, {Ertini},
  {Kuncarayakti}, \& {Andrews}}]{ferrari24}
{Ferrari}, L., {Folatelli}, G., {Ertini}, K., {Kuncarayakti}, H., \& {Andrews},
  J. 2024, arXiv e-prints, arXiv:2406.00130, \dodoi{10.48550/arXiv.2406.00130}

\bibitem[{{Folatelli} {et~al.}(2025){Folatelli}, {Ferrari}, {Ertini},
  {Kuncarayakti}, \& {Maeda}}]{folatelli25}
{Folatelli}, G., {Ferrari}, L., {Ertini}, K., {Kuncarayakti}, H., \& {Maeda},
  K. 2025, \aap, 698, A213, \dodoi{10.1051/0004-6361/202554128}

\bibitem[{{Forde} \& {Goldberg}(2025)}]{forde25}
{Forde}, S., \& {Goldberg}, J.~A. 2025, Research Notes of the American
  Astronomical Society, 9, 135, \dodoi{10.3847/2515-5172/adde46}

\bibitem[{{Glebbeek} {et~al.}(2009){Glebbeek}, {Gaburov}, {de Mink}, {Pols}, \&
  {Portegies Zwart}}]{gleb09}
{Glebbeek}, E., {Gaburov}, E., {de Mink}, S.~E., {Pols}, O.~R., \& {Portegies
  Zwart}, S.~F. 2009, \aap, 497, 255, \dodoi{10.1051/0004-6361/200810425}

\bibitem[{{Goldberg} {et~al.}(2019){Goldberg}, {Bildsten}, \&
  {Paxton}}]{goldberg19}
{Goldberg}, J.~A., {Bildsten}, L., \& {Paxton}, B. 2019, \apj, 879, 3,
  \dodoi{10.3847/1538-4357/ab22b6}

\bibitem[{{Hiramatsu} {et~al.}(2023){Hiramatsu}, {Tsuna}, {Berger}, {Itagaki},
  {Goldberg}, {Gomez}, {Kishalay}, {Hosseinzadeh}, {Bostroem}, {Brown},
  {Arcavi}, {Bieryla}, {Blanchard}, {Esquerdo}, {Farah}, {Howell}, {Matsumoto},
  {McCully}, {Newsome}, {Gonzalez}, {Pellegrino}, {Rhee}, {Terreran},
  {Vink{\'o}}, \& {Wheeler}}]{hiramatsu23}
{Hiramatsu}, D., {Tsuna}, D., {Berger}, E., {et~al.} 2023, \apjl, 955, L8,
  \dodoi{10.3847/2041-8213/acf299}

\bibitem[{{Hsu} {et~al.}(2024){Hsu}, {Smith}, {Goldberg}, {Bostroem},
  {Hosseinzadeh}, {Sand}, {Pearson}, {Hiramatsu}, {Andrews}, {Beasor}, {Dong},
  {Farah}, {Galbany}, {Gomez}, {Padilla Gonzalez}, {Guti{\'e}rrez}, {Howell},
  {K{\"o}nyves-T{\'o}th}, {McCully}, {Newsome}, {Shrestha}, {Terreran},
  {Villar}, \& {Wang}}]{hsu24}
{Hsu}, B., {Smith}, N., {Goldberg}, J.~A., {et~al.} 2024, arXiv e-prints,
  arXiv:2408.07874, \dodoi{10.48550/arXiv.2408.07874}

\bibitem[{{Itagaki}(2023)}]{itagaki23}
{Itagaki}, K. 2023, Transient Name Server Discovery Report, 2023-1158, 1

\bibitem[{{Jacobson-Gal{\'a}n}(2025)}]{jg25}
{Jacobson-Gal{\'a}n}, W.~V. 2025, arXiv e-prints, arXiv:2507.08078,
  \dodoi{10.48550/arXiv.2507.08078}

\bibitem[{{Jencson} {et~al.}(2023){Jencson}, {Pearson}, {Beasor}, {Lau},
  {Andrews}, {Bostroem}, {Dong}, {Engesser}, {Gomez}, {Guolo}, {Hoang},
  {Hosseinzadeh}, {Jha}, {Karambelkar}, {Kasliwal}, {Lundquist}, {Meza
  Retamal}, {Rest}, {Sand}, {Shahbandeh}, {Shrestha}, {Smith}, {Strader},
  {Valenti}, {Wang}, \& {Zenati}}]{jencson23}
{Jencson}, J.~E., {Pearson}, J., {Beasor}, E.~R., {et~al.} 2023, \apjl, 952,
  L30, \dodoi{10.3847/2041-8213/ace618}

\bibitem[{{Jermyn} {et~al.}(2023){Jermyn}, {Bauer}, {Schwab}, {Farmer}, {Ball},
  {Bellinger}, {Dotter}, {Joyce}, {Marchant}, {Mombarg}, {Wolf}, {Sunny Wong},
  {Cinquegrana}, {Farrell}, {Smolec}, {Thoul}, {Cantiello}, {Herwig}, {Toloza},
  {Bildsten}, {Townsend}, \& {Timmes}}]{jermyn23}
{Jermyn}, A.~S., {Bauer}, E.~B., {Schwab}, J., {et~al.} 2023, \apjs, 265, 15,
  \dodoi{10.3847/1538-4365/acae8d}

\bibitem[{{Kasen} \& {Bildsten}(2010)}]{kasen10}
{Kasen}, D., \& {Bildsten}, L. 2010, \apj, 717, 245,
  \dodoi{10.1088/0004-637X/717/1/245}

\bibitem[{{Kilpatrick} {et~al.}(2023){Kilpatrick}, {Foley},
  {Jacobson-Gal{\'a}n}, {Piro}, {Smartt}, {Drout}, {Gagliano}, {Gall},
  {Hjorth}, {Jones}, {Mandel}, {Margutti}, {Ramirez-Ruiz}, {Ransome}, {Villar},
  {Coulter}, {Gao}, {Matthews}, {Taggart}, \& {Zenati}}]{kilpatrick23}
{Kilpatrick}, C.~D., {Foley}, R.~J., {Jacobson-Gal{\'a}n}, W.~V., {et~al.}
  2023, \apjl, 952, L23, \dodoi{10.3847/2041-8213/ace4ca}

\bibitem[{{Kozyreva} {et~al.}(2025){Kozyreva}, {Caputo}, {Baklanov}, {Mironov},
  \& {Janka}}]{kozyreva25}
{Kozyreva}, A., {Caputo}, A., {Baklanov}, P., {Mironov}, A., \& {Janka}, H.-T.
  2025, \aap, 694, A319, \dodoi{10.1051/0004-6361/202452758}

\bibitem[{{Kumar} {et~al.}(2025){Kumar}, {Dastidar}, {Maund}, {Singleton}, \&
  {Sun}}]{kumar25}
{Kumar}, A., {Dastidar}, R., {Maund}, J.~R., {Singleton}, A.~J., \& {Sun},
  N.-C. 2025, \mnras, 538, 659, \dodoi{10.1093/mnras/staf312}

\bibitem[{{Kuriyama} \& {Shigeyama}(2020)}]{kuri20}
{Kuriyama}, N., \& {Shigeyama}, T. 2020, \aap, 635, A127,
  \dodoi{10.1051/0004-6361/201937226}

\bibitem[{{Li} {et~al.}(2025){Li}, {Wang}, {Yang}, {Pastorello}, {Reguitti},
  {Valerin}, {Ochner}, {Cai}, {Iijima}, {Munari}, {Salmaso}, {Farina},
  {Cazzola}, {Trabacchin}, {Fiscale}, {Ciroi}, {Mura}, {Siviero}, {Cabras},
  {Pabst}, {Taubenberger}, {Vogl}, {Fiorin}, {Liu}, {Chen}, {Xiang}, {Mo},
  {Li}, {Wang}, {Zhang}, {Zhai}, {Mirzaqulov}, {Ehgamberdiev}, {Filippenko},
  {Yan}, {Hu}, {Ma}, {Xia}, {Gao}, \& {Li}}]{li25}
{Li}, G., {Wang}, X., {Yang}, Y., {et~al.} 2025, arXiv e-prints,
  arXiv:2504.03856, \dodoi{10.48550/arXiv.2504.03856}

\bibitem[{{Litvinova} \& {Nadezhin}(1985)}]{litvinova85}
{Litvinova}, I.~Y., \& {Nadezhin}, D.~K. 1985, Soviet Astronomy Letters, 11,
  145

\bibitem[{{Liu} {et~al.}(2023){Liu}, {Chen}, {Er}, {Zeimann}, {Vink{\'o}},
  {Wheeler}, {Cooper}, {Davis}, {Farrow}, {Gebhardt}, {Guo}, {Hill}, {House},
  {Kollatschny}, {Kong}, {Kumar}, {Liu}, {Tuttle}, {Endl}, {Duke}, {Cochran},
  {Zhang}, \& {Liu}}]{liu23}
{Liu}, C., {Chen}, X., {Er}, X., {et~al.} 2023, \apjl, 958, L37,
  \dodoi{10.3847/2041-8213/ad0da8}

\bibitem[{{Michel} {et~al.}(2025){Michel}, {Mazzali}, {Perley}, {Hinds}, \&
  {Wise}}]{michel25}
{Michel}, P.~D., {Mazzali}, P.~A., {Perley}, D.~A., {Hinds}, K.~R., \& {Wise},
  J.~L. 2025, \mnras, 539, 633, \dodoi{10.1093/mnras/staf443}

\bibitem[{{Moriya} \& {Singh}(2024)}]{moriya24}
{Moriya}, T.~J., \& {Singh}, A. 2024, arXiv e-prints, arXiv:2406.00928,
  \dodoi{10.48550/arXiv.2406.00928}

\bibitem[{{Morozova} {et~al.}(2016){Morozova}, {Piro}, {Renzo}, \&
  {Ott}}]{morozova16}
{Morozova}, V., {Piro}, A.~L., {Renzo}, M., \& {Ott}, C.~D. 2016, \apj, 829,
  109, \dodoi{10.3847/0004-637X/829/2/109}

\bibitem[{{Morozova} {et~al.}(2015){Morozova}, {Piro}, {Renzo}, {Ott},
  {Clausen}, {Couch}, {Ellis}, \& {Roberts}}]{morozova15}
{Morozova}, V., {Piro}, A.~L., {Renzo}, M., {et~al.} 2015, \apj, 814, 63,
  \dodoi{10.1088/0004-637X/814/1/63}

\bibitem[{{Morozova} \& {Stone}(2018)}]{morozova18}
{Morozova}, V., \& {Stone}, J.~M. 2018, \apj, 867, 4,
  \dodoi{10.3847/1538-4357/aae2b3}

\bibitem[{{Nadyozhin}(2003)}]{nadyozhin03}
{Nadyozhin}, D.~K. 2003, \mnras, 346, 97,
  \dodoi{10.1046/j.1365-2966.2003.07070.x}

\bibitem[{{Nagy} {et~al.}(2014){Nagy}, {Ordasi}, {Vink{\'o}}, \&
  {Wheeler}}]{nagy14}
{Nagy}, A.~P., {Ordasi}, A., {Vink{\'o}}, J., \& {Wheeler}, J.~C. 2014, \aap,
  571, A77, \dodoi{10.1051/0004-6361/201424237}

\bibitem[{{Nagy} \& {Vink{\'o}}(2016)}]{nagy16}
{Nagy}, A.~P., \& {Vink{\'o}}, J. 2016, \aap, 589, A53,
  \dodoi{10.1051/0004-6361/201527931}

\bibitem[{{Neustadt} {et~al.}(2024){Neustadt}, {Kochanek}, \&
  {Smith}}]{neustadt24}
{Neustadt}, J.~M.~M., {Kochanek}, C.~S., \& {Smith}, M.~R. 2024, \mnras, 527,
  5366, \dodoi{10.1093/mnras/stad3073}

\bibitem[{{Niu} {et~al.}(2023){Niu}, {Sun}, {Maund}, {Zhang}, {Zhao}, \&
  {Liu}}]{niu23}
{Niu}, Z., {Sun}, N.-C., {Maund}, J.~R., {et~al.} 2023, \apjl, 955, L15,
  \dodoi{10.3847/2041-8213/acf4e3}

\bibitem[{{Nugis} \& {Lamers}(2000)}]{nugis2000}
{Nugis}, T., \& {Lamers}, H.~J.~G.~L.~M. 2000, \aap, 360, 227

\bibitem[{{Paxton} {et~al.}(2011){Paxton}, {Bildsten}, {Dotter}, {Herwig},
  {Lesaffre}, \& {Timmes}}]{paxton11}
{Paxton}, B., {Bildsten}, L., {Dotter}, A., {et~al.} 2011, \apjs, 192, 3,
  \dodoi{10.1088/0067-0049/192/1/3}

\bibitem[{{Paxton} {et~al.}(2013){Paxton}, {Cantiello}, {Arras}, {Bildsten},
  {Brown}, {Dotter}, {Mankovich}, {Montgomery}, {Stello}, {Timmes}, \&
  {Townsend}}]{paxton13}
{Paxton}, B., {Cantiello}, M., {Arras}, P., {et~al.} 2013, \apjs, 208, 4,
  \dodoi{10.1088/0067-0049/208/1/4}

\bibitem[{{Paxton} {et~al.}(2015){Paxton}, {Marchant}, {Schwab}, {Bauer},
  {Bildsten}, {Cantiello}, {Dessart}, {Farmer}, {Hu}, {Langer}, {Townsend},
  {Townsley}, \& {Timmes}}]{paxton15}
{Paxton}, B., {Marchant}, P., {Schwab}, J., {et~al.} 2015, \apjs, 220, 15,
  \dodoi{10.1088/0067-0049/220/1/15}

\bibitem[{{Paxton} {et~al.}(2018){Paxton}, {Schwab}, {Bauer}, {Bildsten},
  {Blinnikov}, {Duffell}, {Farmer}, {Goldberg}, {Marchant}, {Sorokina},
  {Thoul}, {Townsend}, \& {Timmes}}]{paxton18}
{Paxton}, B., {Schwab}, J., {Bauer}, E.~B., {et~al.} 2018, \apjs, 234, 34,
  \dodoi{10.3847/1538-4365/aaa5a8}

\bibitem[{{Paxton} {et~al.}(2019){Paxton}, {Smolec}, {Schwab}, {Gautschy},
  {Bildsten}, {Cantiello}, {Dotter}, {Farmer}, {Goldberg}, {Jermyn}, {Kanbur},
  {Marchant}, {Thoul}, {Townsend}, {Wolf}, {Zhang}, \& {Timmes}}]{paxton19}
{Paxton}, B., {Smolec}, R., {Schwab}, J., {et~al.} 2019, \apjs, 243, 10,
  \dodoi{10.3847/1538-4365/ab2241}

\bibitem[{{Pledger} \& {Shara}(2023)}]{pledger23}
{Pledger}, J.~L., \& {Shara}, M.~M. 2023, \apjl, 953, L14,
  \dodoi{10.3847/2041-8213/ace88b}

\bibitem[{{Qin} {et~al.}(2023){Qin}, {Zhang}, {Bloom}, {Sollerman},
  {Zimmerman}, {Irani}, {Schulze}, {Gal-Yam}, {Kasliwal}, {Coughlin}, {Perley},
  {Fremling}, \& {Kulkarni}}]{qin23}
{Qin}, Y.-J., {Zhang}, K., {Bloom}, J., {et~al.} 2023, arXiv e-prints,
  arXiv:2309.10022, \dodoi{10.48550/arXiv.2309.10022}

\bibitem[{{Ransome} {et~al.}(2024){Ransome}, {Villar}, {Tartaglia}, {Gonzalez},
  {Jacobson-Gal{\'a}n}, {Kilpatrick}, {Margutti}, {Foley}, {Grayling}, {Ni},
  {Yarza}, {Ye}, {Auchettl}, {de Boer}, {Chambers}, {Coulter}, {Drout},
  {Farias}, {Gall}, {Gao}, {Huber}, {Ibik}, {Jones}, {Khetan}, {Lin},
  {Politsch}, {Raimundo}, {Rest}, {Wainscoat}, {Yadavalli}, \&
  {Zenati}}]{ransome24}
{Ransome}, C.~L., {Villar}, V.~A., {Tartaglia}, A., {et~al.} 2024, \apj, 965,
  93, \dodoi{10.3847/1538-4357/ad2df7}

\bibitem[{{Riess} {et~al.}(2022){Riess}, {Yuan}, {Macri}, {Scolnic}, {Brout},
  {Casertano}, {Jones}, {Murakami}, {Anand}, {Breuval}, {Brink}, {Filippenko},
  {Hoffmann}, {Jha}, {D'arcy Kenworthy}, {Mackenty}, {Stahl}, \&
  {Zheng}}]{riess22}
{Riess}, A.~G., {Yuan}, W., {Macri}, L.~M., {et~al.} 2022, \apjl, 934, L7,
  \dodoi{10.3847/2041-8213/ac5c5b}

\bibitem[{{Shrestha} {et~al.}(2025){Shrestha}, {DeSoto}, {Sand}, {Williams},
  {Hoffman}, {Smith}, {McCall}, {Maund}, {Steele}, {Wiersema}, {Andrews},
  {Smith}, {Bilinski}, {Milne}, {Anche}, {Bostroem}, {Hosseinzadeh}, {Pearson},
  {Leonard}, {Hsu}, {Dong}, {Hoang}, {Janzen}, {Jencson}, {Jha}, {Lundquist},
  {Mehta}, {Retamal}, {Valenti}, {Farah}, {Howell}, {McCully}, {Newsome},
  {Gonzalez}, {Pellegrino}, \& {Terreran}}]{shrestha25}
{Shrestha}, M., {DeSoto}, S., {Sand}, D.~J., {et~al.} 2025, \apjl, 982, L32,
  \dodoi{10.3847/2041-8213/adbb63}

\bibitem[{{Singh} {et~al.}(2024){Singh}, {Teja}, {Moriya}, {Maeda}, {Kawabata},
  {Tanaka}, {Imazawa}, {Nakaoka}, {Gangopadhyay}, {Yamanaka}, {Swain}, {Sahu},
  {Anupama}, {Kumar}, {Anche}, {Sano}, {Raj}, {Agnihotri}, {Bhalerao}, {Bisht},
  {Bisht}, {Belwal}, {Chakrabarti}, {Fujii}, {Nagayama}, {Matsumoto}, {Hamada},
  {Kawabata}, {Kumar}, {Kumar}, {Malkan}, {Smith}, {Sakagami}, {Taguchi},
  {Tominaga}, \& {Watanabe}}]{singh24}
{Singh}, A., {Teja}, R.~S., {Moriya}, T.~J., {et~al.} 2024, arXiv e-prints,
  arXiv:2405.20989, \dodoi{10.48550/arXiv.2405.20989}

\bibitem[{{Soraisam} {et~al.}(2023){Soraisam}, {Matheson}, {Andrews},
  {Narayan}, {Aleo}, \& {ANTARES Team}}]{sorais23}
{Soraisam}, M., {Matheson}, T., {Andrews}, J., {et~al.} 2023, The Astronomer's
  Telegram, 16050, 1

\bibitem[{{Szalai} {et~al.}(2019){Szalai}, {Vink{\'o}}, {K{\"o}nyves-T{\'o}th},
  {Nagy}, {Bostroem}, {S{\'a}rneczky}, {Brown}, {Pejcha}, {B{\'o}di}, {Cseh},
  {Cs{\"o}rnyei}, {Dencs}, {Hanyecz}, {Ign{\'a}cz}, {Kalup}, {Kriskovics},
  {Ordasi}, {P{\'a}l}, {Seli}, {S{\'o}dor}, {Szak{\'a}ts}, {Vida}, {Zsidi},
  {Konkoly Team}, {Arcavi}, {Ashall}, {Burke}, {Galbany}, {Hiramatsu},
  {Hosseinzadeh}, {Hsiao}, {Howell}, {McCully}, {Moran}, {Rho}, {Sand},
  {Shahbandeh}, {Valenti}, {Wang}, {Wheeler}, \& {Supernova
  Project}}]{szalai19}
{Szalai}, T., {Vink{\'o}}, J., {K{\"o}nyves-T{\'o}th}, R., {et~al.} 2019, \apj,
  876, 19, \dodoi{10.3847/1538-4357/ab12d0}

\bibitem[{{Teja} {et~al.}(2023){Teja}, {Singh}, {Basu}, {Anupama}, {Sahu},
  {Dutta}, {Swain}, {Nakaoka}, {Pathak}, {Bhalerao}, {Barway}, {Kumar},
  {A.~J.}, {Imazawa}, {Kumar}, \& {Kawabata}}]{teja23}
{Teja}, R.~S., {Singh}, A., {Basu}, J., {et~al.} 2023, \apjl, 954, L12,
  \dodoi{10.3847/2041-8213/acef20}

\bibitem[{{Tsuna} \& {Takei}(2023)}]{tsuna23}
{Tsuna}, D., \& {Takei}, Y. 2023, \pasj, 75, L19, \dodoi{10.1093/pasj/psad041}

\bibitem[{{Tsuna} {et~al.}(2021){Tsuna}, {Takei}, {Kuriyama}, \&
  {Shigeyama}}]{tsuna21}
{Tsuna}, D., {Takei}, Y., {Kuriyama}, N., \& {Shigeyama}, T. 2021, \pasj, 73,
  1128, \dodoi{10.1093/pasj/psab063}

\bibitem[{{Utrobin} \& {Chugai}(2024)}]{uc24}
{Utrobin}, V.~P., \& {Chugai}, N.~N. 2024, \apss, 369, 49,
  \dodoi{10.1007/s10509-024-04311-9}

\bibitem[{{Utrobin} {et~al.}(2021){Utrobin}, {Chugai}, {Andrews}, {Smith},
  {Jencson}, {Howell}, {Burke}, {Hiramatsu}, {McCully}, \&
  {Bostroem}}]{utrobin21}
{Utrobin}, V.~P., {Chugai}, N.~N., {Andrews}, J.~E., {et~al.} 2021, \mnras,
  505, 116, \dodoi{10.1093/mnras/stab1369}

\bibitem[{{Valenti} {et~al.}(2008){Valenti}, {Benetti}, {Cappellaro}, {Patat},
  {Mazzali}, {Turatto}, {Hurley}, {Maeda}, {Gal-Yam}, {Foley}, {Filippenko},
  {Pastorello}, {Challis}, {Frontera}, {Harutyunyan}, {Iye}, {Kawabata},
  {Kirshner}, {Li}, {Lipkin}, {Matheson}, {Nomoto}, {Ofek}, {Ohyama}, {Pian},
  {Poznanski}, {Salvo}, {Sauer}, {Schmidt}, {Soderberg}, \&
  {Zampieri}}]{valenti08}
{Valenti}, S., {Benetti}, S., {Cappellaro}, E., {et~al.} 2008, \mnras, 383,
  1485, \dodoi{10.1111/j.1365-2966.2007.12647.x}

\bibitem[{{Van Dyk} {et~al.}(2024){Van Dyk}, {Srinivasan}, {Andrews},
  {Soraisam}, {Szalai}, {Howell}, {Isaacson}, {Matheson}, {Petigura},
  {Scicluna}, {Stephens}, {Van Zandt}, {Zheng}, {Chun}, \&
  {Fillippenko}}]{vandyk24}
{Van Dyk}, S.~D., {Srinivasan}, S., {Andrews}, J.~E., {et~al.} 2024, \apj, 968,
  27, \dodoi{10.3847/1538-4357/ad414b}

\bibitem[{{Xiang} {et~al.}(2024){Xiang}, {Mo}, {Wang}, {Wang}, {Zhang}, {Lin},
  \& {Wang}}]{xiang24}
{Xiang}, D., {Mo}, J., {Wang}, L., {et~al.} 2024, Science China Physics,
  Mechanics, and Astronomy, 67, 219514, \dodoi{10.1007/s11433-023-2267-0}

\bibitem[{{Yang} {et~al.}(2024){Yang}, {Liu}, {Pan}, {Er}, {Liu}, {Fang}, {Du},
  {Cai}, {Xu}, {Chen}, {Zou}, {Guo}, {Liu}, {Cheng}, {Kumar}, \&
  {Liu}}]{yang24}
{Yang}, Y.-P., {Liu}, X., {Pan}, Y., {et~al.} 2024, \apj, 969, 126,
  \dodoi{10.3847/1538-4357/ad4be3}

\bibitem[{{Zheng} {et~al.}(2025){Zheng}, {Dessart}, {Filippenko}, {Yang},
  {Brink}, {De Jaeger}, {Vasylyev}, {Van Dyk}, {Patra}, {Jacobson-Galan},
  {Stewart}, {Alvarado}, {Arikatla}, {Beddow}, {Betz}, {Born}, {Bostow},
  {Burgasser}, {Caceres}, {Carrasco}, {Chuang}, {DeGraw}, {Gates},
  {Gendreau-Distler}, {Jacobus}, {Jennings}, {Karpoor}, {Lynam}, {Mina},
  {Mora}, {Pichay}, {Ravi}, {Rees}, {Rich}, {Risin}, {Sandford}, {Savino},
  {Softich}, {Theissen}, {Vidal}, {Wu}, \& {Zeng}}]{zheng25}
{Zheng}, W., {Dessart}, L., {Filippenko}, A.~V., {et~al.} 2025, arXiv e-prints,
  arXiv:2503.13974, \dodoi{10.48550/arXiv.2503.13974}

\bibitem[{{Zimmerman} {et~al.}(2024){Zimmerman}, {Irani}, {Chen}, {Gal-Yam},
  {Schulze}, {Perley}, {Sollerman}, {Filippenko}, {Shenar}, {Yaron}, {Shahaf},
  {Bruch}, {Ofek}, {De Cia}, {Brink}, {Yang}, {Vasylyev}, {Ben Ami}, {Aubert},
  {Badash}, {Bloom}, {Brown}, {De}, {Dimitriadis}, {Fransson}, {Fremling},
  {Hinds}, {Horesh}, {Johansson}, {Kasliwal}, {Kulkarni}, {Kushnir}, {Martin},
  {Matuzewski}, {McGurk}, {Miller}, {Morag}, {Neil}, {Nugent}, {Post},
  {Prusinski}, {Qin}, {Raichoor}, {Riddle}, {Rowe}, {Rusholme}, {Sfaradi},
  {Sjoberg}, {Soumagnac}, {Stein}, {Strotjohann}, {Terwel}, {Wasserman},
  {Wise}, {Wold}, {Yan}, \& {Zhang}}]{zimmerman24}
{Zimmerman}, E.~A., {Irani}, I., {Chen}, P., {et~al.} 2024, \nat, 627, 759,
  \dodoi{10.1038/s41586-024-07116-6}

\end{thebibliography}
\bibliographystyle{aasjournal}

\end{document}